\documentclass[conference,11pt,onecolumn,letterpaper]{IEEEtran}
\usepackage{cite}
\usepackage{wrapfig}
\usepackage[letterpaper, left=1in, right=1in, bottom=1in, top=1in]{geometry}
\usepackage{amsmath,amssymb,amsfonts}
\usepackage{algorithmic}
\usepackage{graphicx}
\usepackage{textcomp}
\usepackage{xcolor}
\usepackage{enumerate}
\usepackage{tikzit}
\input{complexes.tkzs}
\usepackage{amsthm}
\usepackage[utf8]{inputenc} 
\usepackage[T1]{fontenc}    
\usepackage{hyperref}       
\usepackage{url}            
\usepackage{booktabs}       
\usepackage{nicefrac}       
\usepackage{microtype}      

\newtheorem{definition}{Definition}

\usepackage[capitalize]{cleveref}

\Crefname{lemma}{Lemma}{Lemmas}
\Crefname{corollary}{Corollary}{Corollaries}
\usepackage[appendix=append]{apxproof} 
\usepackage[append]{optional}
\renewcommand{\appendixsectionformat}[2]{Additional Details for Section #1 (#2)}

\newtheoremrep{theorem}{Theorem}[section]
\newtheoremrep{lemma}[theorem]{Lemma}
\newtheoremrep{corollary}[theorem]{Corollary}

\newcommand{\I}{\mathcal{I}}
\renewcommand{\O}{\mathcal{O}}
\DeclareMathOperator{\interior}{int}
\DeclareMathOperator{\carr}{carr}
\DeclareMathOperator{\cstar}{star}
\DeclareMathOperator{\ostar}{star^o}
\DeclareMathOperator{\vchi}{|\chi|}
\DeclareMathOperator{\vmu}{|\mu|}
\newcommand{\vsigma}{|\sigma|}
\newcommand{\vI}{|\I|}
\newcommand{\vO}{|\O|}
\newcommand{\vrho}{|\rho|}
\DeclareMathOperator{\Ch}{Ch}
\DeclareMathOperator{\Sub}{Sub}
\DeclareMathOperator{\Bary}{Bary}

\def\BibTeX{{\rm B\kern-.05em{\sc i\kern-.025em b}\kern-.08em
		T\kern-.1667em\lower.7ex\hbox{E}\kern-.125emX}}

\IEEEoverridecommandlockouts
\begin{document}
	
	\title{Continuous Tasks and the Asynchronous Computability Theorem  
\thanks{This work has been supported by the Doctoral College on Resilient Embedded Systems at TU Wien and the Austrian Science Fund (FWF) project ByzDEL (P33600).}
}

	\author{\IEEEauthorblockN{Hugo Rincon Galeana}
		\IEEEauthorblockA{\textit{Embedded Computing Systems Group} \\
			\textit{TU Wien}\\
			Vienna, Austria \\
			0000-0002-8152-1275}
		\and
\IEEEauthorblockN{Sergio Rajsbaum}
		\IEEEauthorblockA{\textit{Instituto de Matemáticas } \\
			\textit{UNAM}\\
			Mexico City, Mexico \\
			0000-0002-0009-5287}
		\and
		\IEEEauthorblockN{Ulrich Schmid}
		\IEEEauthorblockA{\textit{Embedded Computing Systems Group} \\
			\textit{TU Wien}\\
			Vienna, Austria \\
			0000-0001-9831-8583}
	}
	
	\maketitle
	
	\begin{abstract}
The celebrated 1999 Asynchronous Computability Theorem (ACT)
of Herlihy and Shavit characterized distributed tasks
that are wait-free solvable and uncovered deep connections with 
combinatorial topology. We provide an alternative characterization of
those tasks by means of the novel concept of continuous tasks, 
which have an input/output specification that is a continuous function between 
the geometric realizations of the input and output complex: We state and
prove a precise characterization theorem (CACT) for wait-free solvable tasks in 
terms of continuous tasks. Its proof utilizes a novel chromatic 
version of a foundational result in algebraic topology, 
the simplicial approximation theorem, which is also proved in this paper.
Apart from the simple alternative proof of the ACT implied by our CACT,
we also demonstrate that continuous tasks have an expressive power that 
goes way beyond classic task specifications, and hence open up a promising 
venue for future research: 
For the well-known approximate agreement task, we show that 
one can easily encode the desired density of the occurrence of specific 
outputs, namely, exact agreement, in the continuous task specification.
	\end{abstract}
	
	\begin{IEEEkeywords}
		Wait-free computability, topology, distributed computing, decision tasks, shared memory.
		\end{IEEEkeywords}
		

\section{Introduction}
\label{sec:intro}

Given a finite set of values $V$, a \emph{task} $\langle \mathcal{I},\mathcal{O},\Delta\rangle$ is a problem where each process of a distributed system starts with a
private input value from $V$, communicates with the others, and halts with a private
output value from $V$. 
The input complex $\mathcal{I}$ defines the set of possible assignments of input values to the processes, and the output
complex $\mathcal{O}$ defines the allowed decisions. The input/output relation  $\Delta$ 
specifies, for each input assignment $\sigma\in\mathcal{I}$, a set $\Delta(\sigma)\subseteq \mathcal{O}$ 
of valid output decisions. 
Processes are usually \emph{asynchronous}, they can be halted or
delayed without warning by cache misses, interrupts, or scheduler pre-emption.
 In asynchronous
systems, it is desirable to design algorithms that are \emph{wait-free}: any process that
continues to run will produce an output value,
regardless of delays or failures by other processes.

It is now almost 30 years since Herlihy and Shavit~\cite{HS99:ACT} presented the celebrated \emph{Asynchronous Computability Theorem} (ACT).
Roughly speaking, the theorem states that a task $\langle \mathcal{I},\mathcal{O},\Delta\rangle$
is wait-free solvable in a
read/write shared memory distributed system if and only if  $\mathcal{I}$
can be subdivided $r$ times, and sent by a simplicial decision map $\mu$
to $\mathcal{O}$, respecting the input/output relation (carrier map) $\Delta$. 
Intuitively, $r$ is the number of rounds  the processes
need to communicate to solve the task, and hence yields a time complexity characterization
as well~\cite{HoestS06}.

\cref{th:ACTif} below states one direction of the ACT (restricted for reference purposes only); it actually holds
in the other direction as well. 
It is the easier direction, and can be proved in several ways, see e.g.~\cite{AttiyaR02,BG93:STOC,HS99:ACT,SZ00:SIAM},
 essentially  by considering an appropriate subset
of all wait-free executions of a full-information protocol:
 in any wait-free read/write model,
any protocol  solving a given task induces a subdivision of the task's input complex.

\begin{theorem}[ACT (if-direction)~\cite{HS99:ACT}]
\label{th:ACTif}
If $\langle \mathcal{I}, \mathcal{O}, \Delta \rangle$ is solvable, then there exists a chromatic subdivision $\Sub(\mathcal{I})$ and a chromatic simplicial map $\mu : \Sub(\mathcal{I}) \rightarrow \mathcal{O}$ carried by $\Delta$. 
\end{theorem}

The  ACT  reinterprets distributed computing geometrically,
and provides an explanation of why some tasks are solvable, and not others.
It is particularly useful for proving that some tasks are unsolvable, opening the doors
to the very powerful machinery of combinatorial topology; notable examples
are  the set agreement~\cite{HS99:ACT} impossibility result, which generalizes the classic FLP consensus
impossibility~\cite{FLP85}, 
and the
renaming~\cite{Castaeda2010NewCT} impossibility result.
Furthermore, the theorem is the basis for task solvability characterizations
in other distributed computing models: systems where at most $t$ processes may crash, synchronous and partially synchronous processes,  Byzantine and dependent failures, 
stronger shared memory communication objects,  message passing models and even robot coordination algorithms~\cite{AlcantaraCFR19}, see \cref{sec:relatedWork}. 
An overview of topological distributed computing theory, as started by the ACT, can be found in the book~\cite{HKR13}.

\medskip

The simplicial formulation of the input/output relation $\Delta$ and the ACT
in general both obfuscates certain properties and introduces technical 
difficulties, which considerably complicated the proofs, 
cp.~\cite{AttiyaR02,BG93:STOC,HS99:ACT,SZ00:SIAM}.
For \emph{colorless} tasks like set agreement, which can be defined
without refering to process ids (discussed in more detail
in \cref{sec:relatedWork}), there is also 
a \emph{continuous} version of the ACT. It roughly
says that a colorless task is wait-free solvable
if and only if  there is a continuous map $f$ from $|\mathcal{I}|$ to  
  $|\mathcal{O}|$ carried by  $\Delta$, where $|\mathcal{K}|$ denotes 
the geometric realization of the complex $\mathcal{K}$. Since all
involved complexes are colorless here, the proofs are much simpler.
However, we are
not aware of any formulation of the ACT via continuous functions for 
general (chromatic) tasks. 

Indeed, the main source of the technical difficulties in the proof of the ACT 
is that the objects involved are \emph{chromatic}: each vertex of $\mathcal{I}$
and $\mathcal{O}$ is
associated to one of the process ids in the system, and all the
simplicial maps are required to preserve vertex ids, i.e., must be
\emph{rigid} in that they always preserves the dimension of the
simplices. Quite some effort is needed to deal with the resulting
requirement of a chromatic decision map $\mu$ in the case of
a (non-colorless) task such as renaming, which can specify which 
values can be output by which process. 

 \begin{wrapfigure}[7]{r}{0.4\textwidth}
 \centering
 \vspace{-0.9cm}
\scalebox{0.6}{\begin{tikzpicture}
	\begin{pgfonlayer}{nodelayer}
		\node [style=VG] (0) at (-16, 0) {};
		\node [style=VR] (1) at (-9, 0) {};
		\node [style=VY] (2) at (-12.5, 5) {};
		\node [style=none] (11) at (-16, 0) {};
		\node [style=none] (12) at (-12.5, 5) {};
		\node [style=none] (13) at (-9, 0) {};
		\node [style=none] (22) at (-12.75, -1.75) {$\I$};
		\node [style=none] (34) at (-1.5, -1.75) {$\O$};
		\node [style=VG] (35) at (-5, 0) {0};
		\node [style=VR] (36) at (2, 0) {0};
		\node [style=VY] (37) at (-1.5, 3) {1};
		\node [style=none] (38) at (-5, 0) {};
		\node [style=none] (39) at (-1.5, 3) {};
		\node [style=none] (40) at (2, 0) {};
		\node [style=VG] (41) at (-0.5, 0) {2};
		\node [style=VG] (42) at (-3, 5) {1};
		\node [style=VR] (43) at (0, 5) {1};
		\node [style=VY] (44) at (-1.5, 7) {0};
		\node [style=none] (45) at (-3, 5) {};
		\node [style=none] (46) at (-1.5, 7) {};
		\node [style=none] (47) at (0, 5) {};
		\node [style=none] (48) at (-1.5, 3) {};
		\node [style=VR] (49) at (-2.5, 0) {2};
		\node [style=none] (50) at (-14.5, 3.5) {};
		\node [style=none] (51) at (-3, 6) {};
		\node [style=none] (52) at (-8.25, 6.75) {$\Delta$};
	\end{pgfonlayer}
	\begin{pgfonlayer}{edgelayer}
		\draw [style=EdgeRed] (0) to (2);
		\draw (2) to (1);
		\draw (1) to (0);
		\draw [style=FaceBlue] (12.center)
			 to [in=120, out=-45, looseness=0.00] (13.center)
			 to (11.center)
			 to cycle;
		\draw [style=EdgeRed] (35) to (37);
		\draw (37) to (36);
		\draw (36) to (35);
		\draw [style=FaceBlue] (39.center)
			 to [in=120, out=-45, looseness=0.00] (40.center)
			 to (38.center)
			 to cycle;
		\draw (43) to (42);
		\draw [style=FaceBlue] (48.center)
			 to (47.center)
			 to (46.center)
			 to (45.center)
			 to cycle;
		\draw (48.center) to (47.center);
		\draw (47.center) to (46.center);
		\draw [style=EdgeRed] (46.center) to (45.center);
		\draw [style=EdgeRed] (45.center) to (48.center);
		\draw (49) to (48.center);
		\draw (41) to (48.center);
		\draw [style=Eout, bend left=60] (50.center) to (51.center);
	\end{pgfonlayer}
\end{tikzpicture}}
\end{wrapfigure} 
This is illustrated by the \emph{Hourglass task} shown to the right, where there are three processes denoted by green, red, and yellow. There is no input, hence only one input simplex, and the processes must output labels in $\{0,1,2\}$ as shown in the output complex. The carrier map $\Delta$ defining this task 
requires 
(i) to map the corners (solo executions)
of $\mathcal{I}$ to  the vertices labeled $0$ in $\mathcal{O}$,
(ii) to map the boundary (executions where two processes participate)
of $\mathcal{I}$ to the boundary of $\mathcal{O}$, as shown for the yellow-green boundary
in the figure; and 
(iii) the triangle of $\mathcal{I}$ (executions where all participate), to any triangle in $\mathcal{O}$.

Interestingly, there is a continuous map $f$ from $|\mathcal{I}|$ to  
  $|\mathcal{O}|$ carried by  $\Delta$, i.e., the Hourglass task would
satisfy the conditions of the colorless ACT. It fails to meet our conditions 
for a continuous task (see~\cref{def:contas}), however. And indeed, it has already been proved 
in~\cite{HKR13} that it is not wait-free solvable. By contrast, as 
the output complex of the Hourglass trask is not link-connected, 
it is not possible to apply related characterization theorems 
like \cite[Thm.~8.4]{GKM14:PODC} to prove this unsolvability (see 
\cref{sec:relatedWork} for a more detailed discussion).

 %

%

\medskip

A foundational result in algebraic topology is the \emph{simplicial approximation theorem}  first proved by Brouwer, which
served to put  homology theory on  a rigorous basis~\cite{simpComp}.
It guarantees that continuous maps can be (by a slight deformation) approximated by simplicial
maps.
Roughly, it says that given a continuous map $f$  from $|\mathcal{I}|$ to  
  $|\mathcal{O}|$, there is some $r$ such that $f$ has a corresponding simplicial map $\mu$
  from $\Bary^r( \mathcal{I})$ to $\mathcal{O}$ that is a simplicial approximation, where $\Bary$ denotes the barycentric subdivision (or any other mesh-shrinking subdivision).
The colorless version of the ACT \cite[Ch.~4]{HKR13} is essentially a distributed computing version of the simplicial approximation theorem. And indeed, in the case of the Hourglass task, the simplicial approximation theorem
says that there is some $r$ such that there is a simplicial map $\mu$
  from $\Bary^r( \mathcal{I})$ to $\mathcal{O}$ respecting $\Delta$. However, 
  this does not result in a wait-free algorithm, since there is no such $\mu$ 
that preserves colors, i.e., is rigid. 

Whereas several constructions of
a chromatic simplicial approximation have been used in the existing proofs
of the ACT \cite{HS99:ACT,AttiyaR02,BG93:STOC,SZ00:SIAM} and some 
generalizations \cite{GKM14:PODC,SaraphHG18}, they are complicated
and tailored to the specific context, which ensures link connectivity. 
We are not aware of a simplicial approximation theorem that could guarantee 
a chromatic map $\mu$ under more general general conditions, like the
ones for our continuous tasks (see~\cref{def:contas}).
  
\medskip

\noindent
\textbf{Contributions:} Our paper has three main contributions:
(1) we introduce the novel notion of a continuous task and show that it allows for  more expressive task specifications,
(2) we formulate and prove a chromatic version of the simplicial approximation theorem, and (3) we use these notions to formulate a continuous version the ACT, 
denoted CACT, which we prove to be equivalent to the ACT in the wait-free  shared memory model. 
Rather than focusing on a specific model of computation, however, the only property we require from the model is that \cref{th:ACTif} holds, i.e., if some protocol solves a task, then the protocol determines a chromatic subdivision of the input complex.
We refer to any such model 
by ASM, and give examples of such models (such as the read-write shared memory model of the original ACT~\cite{HS99:ACT} and the \emph{Iterated Immediate Snapshot} (IIS) model) in~\cref{sec:relatedWork}.

In more detail, our paper contains the following contributions:
\begin{enumerate}
\item[(1)] We introduce a \emph{continuous task} as a triple
$\langle \mathcal{I},\mathcal{O},f\rangle$: The possible input and output configurations
are determined by the chromatic input and output complexes $\mathcal{I},\mathcal{O}$, as in the usual
task notion.
The input/output specification $f$ is a continuous function from  
$|\mathcal{I}|$ to $|\mathcal{O}|$, instead of an input/output relation, $\Delta$.
We identify an additional property called 
\emph{chromatic} for $f$, which intuitively requires $f$ to satisfy a minimal color  and local
dimension preserving requirement, stated formally in \cref{def:chrfun}.

Semantically, continuous tasks  present interesting expressive facilities with respect to
traditional input/output specifications in the form of a carrier map $\Delta$. 
Indeed, they open up various interesting research
questions, which we can barely touch here: In \cref{sec:app}, we introduce refined versions of the well-known $1/3$-approximate agreement task, which demonstrate that even constraints on the density of the outputs for a given input can be expressed by means of continuous task specifications.


\item[(2)] We state and prove \cref{thm:chrappr}, a chromatic version of the simplicial approximation theorem, for chromatic functions (see~\cref{def:chrfun}), which may be of independent interest also. In a way, it off-loads part of the complexity of constructing geometric subdivisions that ensure rigid maps for an \emph{arbitrary} continuous function $f$ to the definition of a chromatic function $f$.

\item[(3)] Using our chromatic simplicial approximation  \cref{thm:chrappr}, we prove that chromatic functions  precisely capture the notion of  solvability in an ASM model. This leads to our CACT \cref{thm:asmeq}, which states that a task 
$\langle \mathcal{I}, \mathcal{O}, \Delta \rangle$ is wait-free solvable  if and only if there exists a corresponding continuous task $\langle \mathcal{I}, \mathcal{O}, f \rangle$. Finally, \cref{cor:main} states that any continuous task is solvable, which implies the converse of \cref{th:ACTif}.

\end{enumerate}


Overall, our results provide a refined explanation on
the reasons of why a task may or may not be solvable in an ASM model, and provide
an alternative perspective and proof of the ACT (\cref{sec:relatedWork} discusses  several related theorems).

\medskip

\noindent
\textbf{Paper organization:}
In \cref{sec:cont}, we define continuous tasks and the meaning of solving such a task, and present the chromatic simplicial approximation theorem.
\cref{sec:results} contains our CACT theorem, and \cref{sec:app} is devoted to
the application showing the expressive power of continuous tasks. A discussion of additional 
related work and some conclusions and directions of future research are provided in~\cref{sec:relatedWork} and \cref{sec:concl}, respectively.
In the appendix,  we provide and all our proofs as well as a collection of background combinatorial topology and distributed computing definitions, primarily taken from~\cite{HKR13}.


\section{Continuous Tasks and the Chromatic Simplicial Approximation Theorem}
\label{sec:cont}

In this section, we extend the standard language (see e.g.\ \cite{HKR13}) 
of combinatorial topology in distributed computing to be able to specify continuous tasks. More specifically, given arbitrary geometric realizations $|\mathcal{I}|$ and $|\mathcal{O}|$, i.e., metric topological spaces formed as the union of geometric simplices $|\sigma|$ corresponding to the respective abstract simplices in $\mathcal{I}$ and $\mathcal{O}$, we introduce the concept of a chromatic function, which is a continuous function that plays the role of a minimal carrier map that determines task solvability. 

\subsection{Chromatic Functions and Continuous Task Solvability}
\label{sec:chFunct}

We first need to extend the notion of a coloring in order to assign a set of colors to a point in 
the geometric realization $|\mathcal{K}|$ of a finite simplicial complex $\mathcal{K}$.
We start with the carrier $\textrm{carr}(x, \mathcal{K})$ of a point $x$ in the geometric realization $\vert \mathcal{K} \vert$ of a simplicial complex $\mathcal{K}$.

\begin{definition}\label{def:carrgeo}
 For a point $x \in \vert \mathcal{K} \vert$, let the carrier of $x$, denoted $\sigma = \textrm{carr}(x, \mathcal{K})$, be the unique smallest simplex $\sigma \in \mathcal{K}$ such that $x \in \vert \sigma \vert$. We can also define the carrier of a set $S \subseteq \vert \mathcal{K} \vert, \textrm{ as } \textrm{carr}(S,\mathcal{K}) = \displaystyle \bigcup\limits_{x \in S} \textrm{carr}(x, \mathcal{K})$.
\end{definition}

\begin{definition}[Extended coloring]\label{def:extchr}
Let $\mathcal{K}$ be an $m$-dimensional simplicial complex with a coloring $\mathcal{\chi}: \mathcal{K} \rightarrow 2^\Pi = 2^{\{1, \ldots, m \}}$. We define the extended coloring $\vert \chi\vert: \vert \mathcal{K} \vert \rightarrow 2^\Pi$ with respect to $\mathcal{K}$  as $\vert \chi \vert(x) = \chi(\textrm{carr}(x, \mathcal{K}))$ for every $x \in \vert \mathcal{K} \vert $.
\end{definition}

\begin{figure}[h!]
	\centering
\scalebox{0.7}{\begin{tikzpicture}
	\begin{pgfonlayer}{nodelayer}
		\node [style=VG] (0) at (-16, 0) {};
		\node [style=VR] (1) at (-9, 0) {};
		\node [style=VY] (2) at (-12.5, 5) {};
		\node [style=none] (11) at (-12.5, 5) {};
		\node [style=none] (12) at (-16, 0) {};
		\node [style=none] (13) at (-9, 0) {};
		\node [style=none] (22) at (-12.75, -1.75) {$\vI$};
		\node [style=none] (23) at (-12.5, 1.75) {};
		\node [style=none] (24) at (-12.5, 1.75) {};
		\node [style=none] (25) at (-12.5, 1.75) {};
		\node [style=none] (26) at (-12.5, 1.25) {$x$};
		\node [style=none] (27) at (-12.5, 3.5) {};
		\node [style=none] (28) at (-13.25, 2) {};
		\node [style=none] (29) at (-11.5, 1.75) {};
		\node [style=VBsmall] (30) at (-12.5, 2.5) {};
		\node [style=none] (31) at (-13.75, 2.75) {};
		\node [style=none] (32) at (-11.25, 2.25) {};
	\end{pgfonlayer}
	\begin{pgfonlayer}{edgelayer}
		\draw (0) to (2);
		\draw (2) to (1);
		\draw (1) to (0);
		\draw [style=FaceBlue] (12.center)
			 to [in=120, out=-45, looseness=0.00] (13.center)
			 to (11.center)
			 to cycle;
		\draw [style=FaceYellow] (28.center)
			 to (29.center)
			 to (27.center)
			 to cycle;
		\draw [style=EdgeRed] (31.center) to (32.center);
	\end{pgfonlayer}
\end{tikzpicture}}
	\caption{The inner (yellow) triangle, and the red segment are neighborhoods of $x\in\interior(\vsigma)$ of dimension $2$ and $1$, respectively, in the outer (blue) triangle $\vsigma$.}
	\label{fig:neighborhood}
\end{figure}
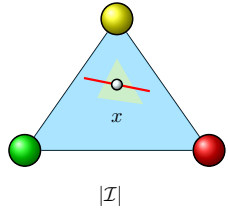

In addition to defining an extended coloring, we also need a notion of closeness. In topology, a neighborhood of a point $x$ is a collection of open sets that include $x$ and defines what does it mean to be close to $x$. We will use the following \cref{def:simpneigh}, illustrated in \cref{fig:neighborhood}.

\begin{definition}[Simplicial neighborhood]\label{def:simpneigh}
	Let $\mathcal{I}$ be an abstract simplicial complex, $x \in \vI$ a point in its geometric realization, and $\sigma = \carr(x, \I)$. We say that $N \subseteq \vsigma$ is a simplicial neighborhood of $x$ if $ N \cong \vert \{x_1, x_2, \ldots, x_r\} \vert$, where each $x_i \in \vsigma$, and $x \in \interior(N)$ is a point in its interior. That is, $N$ is homeomorphic to a geometric simplex generated from points that belong to $\vsigma$.
\end{definition}

A chromatic function has the property that it does not allow to map a neighborhood of $x$ to a neighborhood of $f(x)$ with smaller dimension, and which preserves colors:

\begin{definition}[Chromatic Function]\label{def:chrfun} We say that a continuous function $f: \vert \mathcal{I} \vert \rightarrow \vert \mathcal{O} \vert $ is chromatic (with respect to $\mathcal{I}$ and $\mathcal{O}$) if, for every $x \in \vert \mathcal{I} \vert $, and any simplicial neighborhood $N$ of $x$, it guarantees $\vchi(N) \cap \vchi(f(N)) \geq \dim(N)+1$.
\end{definition}

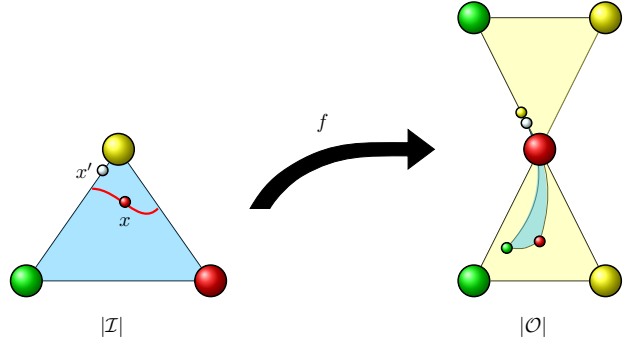
\begin{figure}[h!]
\centering
\scalebox{0.7}{\begin{tikzpicture}
	\begin{pgfonlayer}{nodelayer}
		\node [style=VG] (0) at (-16, 0) {};
		\node [style=VR] (1) at (-9, 0) {};
		\node [style=VY] (2) at (-12.5, 5) {};
		\node [style=VG] (3) at (1, 0) {};
		\node [style=VY] (4) at (6, 0) {};
		\node [style=VR] (5) at (3.5, 5) {};
		\node [style=VG] (6) at (1, 10) {};
		\node [style=VY] (7) at (6, 10) {};
		\node [style=VYsmall] (8a) at (2.8, 6.4) {};
		\node [style=VBsmall] (8) at (3.0, 6.0) {};
		\node [style=VRsmall] (9) at (3.5, 1.5) {};
		\node [style=VGsmall] (10) at (2.25, 1.25) {};
		\node [style=none] (11) at (-16, 0) {};
		\node [style=none] (12) at (-12.5, 5) {};
		\node [style=none] (13) at (-9, 0) {};
		\node [style=none] (14) at (1, 10) {};
		\node [style=none] (15) at (6, 10) {};
		\node [style=none] (16) at (3.5, 1.5) {};
		\node [style=none] (17) at (2.25, 1.25) {};
		\node [style=none] (18) at (3.5, 7.5) {};
		\node [style=none] (19) at (3.5, 5) {};
		\node [style=none] (20) at (6, 0) {};
		\node [style=none] (21) at (1, 0) {};
		\node [style=none] (22) at (-12.75, -1.75) {$\vI$};
		\node [style=none] (23) at (3.25, -1.75) {$\vO$};
		\node [style=none] (26) at (-7.5, 2.75) {};
		\node [style=none] (27) at (-5.75, 4.5) {};
		\node [style=none] (28) at (-1.5, 5.25) {};
		\node [style=none] (29) at (-6.75, 2.75) {};
		\node [style=none] (30) at (-5.5, 4) {};
		\node [style=none] (31) at (-1.5, 4.75) {};
		\node [style=none] (32) at (-1.5, 5.75) {};
		\node [style=none] (33) at (-1.5, 4.25) {};
		\node [style=none] (34) at (-0.5, 5) {};
		\node [style=none] (35) at (-4.75, 6) {$f$};
		\node [style=none] (36) at (-13.5, 3.5) {};
		\node [style=none] (36b) at (-13.8, 4.2) {$x'$};
		\node [style=VBsmall] (36c) at (-13.1, 4.2) {};
		\node [style=none] (37) at (-11, 2.75) {};
		\node [style=none] (38) at (-12.25, 2.25) {$x$};
		\node [style=VRsmall] (39) at (-12.25, 3) {};
	\end{pgfonlayer}
	\begin{pgfonlayer}{edgelayer}
		\draw (0) to (2);
		\draw (2) to (1);
		\draw (1) to (0);
		\draw (3) to (4);
		\draw (4) to (5);
		\draw (3) to (5);
		\draw (5) to (6);
		\draw (5) to (7);
		\draw (7) to (6);
		\draw [bend right=328] (8) to (10);
		\draw [bend right=15, looseness=1.25] (10) to (9);
		\draw [bend right=338] (8) to (9);
		\draw [style=FaceBlue] (12.center)
			 to [in=120, out=-45, looseness=0.00] (13.center)
			 to (11.center)
			 to cycle;
		\draw [style=FaceYellow] (20.center)
			 to (21.center)
			 to (19.center)
			 to cycle;
		\draw [style=FaceYellow] (19.center) to (15.center);
		\draw [style=FaceYellow] (19.center)
			 to (15.center)
			 to (14.center)
			 to cycle;
		\draw [style=FaceBlue] (17.center)
			 to [bend right=15, looseness=1.25] (16.center)
			 to [bend right=22] (8.center)
			 to [bend left=32] cycle;
		\draw [style=FaceYellow, in=-135, out=135, loop] (17.center) to ();
		\draw [style=ArrowBlack] (34.center)
			 to (32.center)
			 to [in=90, out=-90] (28.center)
			 to [in=30, out=180] (27.center)
			 to [bend right=15] (26.center)
			 to (29.center)
			 to [bend left=15, looseness=0.75] (30.center)
			 to [in=180, out=30] (31.center)
			 to [in=90, out=-90] (33.center)
			 to cycle;
		\draw [style=EdgeRed, in=-135, out=0] (36.center) to (37.center);
	\end{pgfonlayer}
\end{tikzpicture}}
	\caption{$f$ is non-chromatic, since it maps a 1-dimensional neighborhood of a point $x$ (the red curve) with $\carr(x,\I)=\{\textrm{red}, \textrm{green}, \textrm{yellow}\}$ to the 0-dimensional red central vertex. It also violates the color preservation
requirement, as a 1-dimensional neighborhood of $x'$ 
within the (green, yellow) edge in $\vI$ is mapped onto the (red, green) 
edge in the upper simplex in $\vO$.}
	\label{fig:nonchromatic}
\end{figure}

We note that not every continuous function is chromatic. An example is shown
in \cref{fig:nonchromatic}, where we assume that $f$ maps the entire red
curve in the blue simplex $\vsigma \in \vI$ on the left to the central red 
vertex in $\vO$ on the right. This collapses a 1-dimensional neighborhood $N$ 
of $x$ to a 0-dimensional neighborhood $f(N)$. Similarly, if we assume
that $f$ also maps the whole area above this curve to the line connecting 
the central red vertex to the small yellow vertex on the boundary of the 
upper simplex 
in $\vO$, it also collapses a 2-dimensional neighborhood to a 
1-dimensional neighborhood. Finally, $f$ also violates the color preservation
requirement, as a 1-dimensional neighborhood $N'\in\vrho$ of $x'$ lying 
in the face $\vrho \in \vsigma$ consisting 
of the (green, yellow) edge is mapped to $f(N')$, which lies on the
the (red, green) edge in the upper simplex in $\vO$.

\cref{lemma:subchr} and \cref{cor:subchr} show that it is possible to chromatically subdivide 
the input complex while mantaining the color preservation property of a given chromatic function $f$. Indeed, even if some chromatic subdivision does not fit w.r.t.\ color preservation, we can make an arbitrarily small perturbation to the vertices in the subdivision fitting. We note that the proof of this lemma is substantially less involved than the perturbation argument in the proof of the  ACT~\cite{HS99:ACT}, \cite[Ch.~11]{HKR13}. In fact, rather than constructing subdivisions that ensure rigidity for arbitrary continuous functions $f$, we only need subdivisions that allow $f$ to remain a chromatic function.

\begin{lemmarep} \label{lemma:subchr}
	Let $f: \vert \mathcal{I} \vert  \rightarrow \vert \mathcal{O} \vert$ be a chromatic function with respect to $\mathcal{I}$ and $\mathcal{O}$. For any 1-layer chromatic subdivision $\Ch(\I)$ of $I$, there exist a geometric realization $|\Ch(\I)|$ such that $f: |\Ch(\I)| \rightarrow \vO$ is chromatic with respect to $\Ch(\I)$ and $\mathcal{O}$.
\end{lemmarep}

\begin{proof}
		To prove the lemma, it suffices to show that $f$ is chromatic for any given face $\Ch(\sigma)$ of dimension $k$. For $k=0$, this is trivial, so assume that we have shown this already for all faces of dimension $k-1$, and consider a face $\Ch(\sigma)$ with $\dim(\I) = k\geq 1$. Let $x\in \interior(\vsigma)$ be an internal point of $|\sigma|$.  Since $f$ is chromatic with respect to $\I$, there exists a $k$-dimensional simplicial neighborhood $N\subseteq \interior(\vsigma)$ of $x$ such that $|\chi|(f(N)) = \{1, \ldots, k+1\}$, i.e., includes all colors. We will first show that there is a point $y \in f(N)$ such that $|\chi|(y) = \{1, \ldots, k+1\}$: Assuming the contrary, $f(N)$ would be contained in the $k-1$ skeleton of $\vO$. Consequently, also $f(x)$ does not include all colors, i.e., there exists a color $i \notin |\chi| (f(x))$. Since the $k-1$ skeleton of $\vO$ that includes color $i$ is a closed set that does not include $f(x)$, there is a $k$-dimentional $\varepsilon$-ball $B_{\varepsilon}(f(x))$ around $f(x)$ that does not intersect the $k-1$ skeleton of $\vO$ that includes $i$ as a color. Due to continuity of $f$, there exists a $k$-dimensional $\delta$-ball $B_\delta(x)$ around $x$ that is contained in $N$, which is mapped to  $B_{\varepsilon}(f(x))$. Since every point in $N\subseteq \interior(\vsigma)$ has all $k+1$ colors in its carrier, we also obtain $|\chi| (B_\delta(x))  = k+1$, but $| \chi | (f(B_\delta(x))) \leq k$, since it does not include $i$. This contradicts that $f$ is chromatic. 
	
	Therefore, there exists a point $y \in f(N)$ such that $|\chi|(y) = \{1, \ldots, k+1\}$. Since $y$ is hence an interior point in $\vO$, there exists $B_{\varepsilon}(y) \subseteq \vO$ where every point has all colors in its carrier. Since $y \in f(N)$ and $N\subseteq \interior(\vsigma)$, there exists $z \in \interior(\vsigma)$ such that $f(z) = y$. Again from continuity of $f$, it follows that there exists a $\delta$-ball $B_{\delta}(z)$ around $z$ that maps to $f(B_\delta(z)) \subseteq B_\epsilon(y)$, where every point in the latter has all colors in its carrier. Therefore, in the to-be-constructed subdivision $\Ch(\sigma)$, we can create the new central face $\sigma'$ of dimension $k$ inside $B_{\delta}(z)$, which will make $f$ chromatic with respect to $|\sigma'|$.
	
According to the induction hypothesis, we have already subdivided the boundary of $\sigma$ (with dimension $< k$) in a way that makes $f$ chromatic.
What still remains to be done, however, is to show that $f$ is also chromatic in the non-central faces of dimension $k$ generated by our subdivision. For this purpose, it suffices to show that every $k-1$-dimensional face $\rho'$ that includes at least one vertex from $\sigma'$ can be chosen such that it includes all its $k$ colors $\{1,\ldots,k\}$. This suffices, since lower-dimensional faces
of $\rho'$ inherently preserve the chromatic property of $f$: (1) For lower-dimensional faces on the boundary of $\sigma$, this follows from the induction hypothesis. (2) For $k-2$-dimensional faces in the interior of $\sigma$ that originate from the intersection $\rho' \cap \rho''$ with another $k-1$-dimensional face $\rho''$, the intersection of their color sets must contain $k-2$ colors, which guarantees that $f$ is chromatic also on $\rho' \cap \rho''$. Finally, (3) for lower-dimensional faces lying in $\sigma'$, the color set even comprises all $k$ colors and cannot hence make $f$ non-chromatic.

To finally justify that, for any $k-1$-dimensional face $\rho'=(v_1, \ldots ,v_{k})$ of vertices in our subdivision, there must indeed be a simplicial neighborhood $N_{k-1}'$ that connects these vertices in $N$ such that $|\chi| (N_{k-1}')=\{1,\ldots,k\}$, we just repeat the argument in our first step above: Assuming the contrary, $f(N_{k-1}')$ would be in the $k-2$-skeleton of $\vO$, which does not include $\{1,\ldots,k\}$ and would hence lead to a contradiction to $f$ being chromatic with respect to $\I$ and $\O$. Consequently, we can always find a suitable choice for the geometric realization for $\rho'$, namely $N_{k-1}'$. According to case (2) above, for neighboring faces $\rho'$ and $\rho''$,
we can choose the common border $\rho'\cap\rho''$ arbitrarily within $N_{k-1}' \cap N_{k-1}''$.
	
This completes the proof, since we showed that we can insert interior faces and connect new vertices
in the geometric subdivision without making $f$ non-chromatic.
\end{proof}

\begin{corollary}\label{cor:subchr}
	Let $f: \vert \mathcal{I} \vert \rightarrow \vert \mathcal{O} \vert$ be a chromatic function with respect to $\mathcal{I}$ and $\mathcal{O}$. For any chromatic subdivision $\Sub(\I)$ of 
$\mathcal{I}$, there exists a geometric realization $|\Sub(\I)|$ of such that $f: |\Sub(\I)| \rightarrow \vO$ is chromatic with respect to $\Sub(\I)$ and $\mathcal{O}$.
\end{corollary}

Our first step towards establishing a discrete/continuous duality between chromatic functions and chromatic simplicial maps, i.e., simplicial maps that preserve the colors of vertices (and hence are rigid), is showing that the geometric realization of a chromatic simplicial map is a chromatic function. 

\begin{lemmarep}\label{lemma:chrmapcont}
	Let $\mu : \Ch^k(\mathcal{I}) \rightarrow  \mathcal{O}$ be a chromatic simplicial map. Then $\vmu: \vI \rightarrow \vO$ is a chromatic function.
\end{lemmarep}

\begin{proof}
	Continuity follows from $\vmu$ being an affine mapping from $\vert \Ch^k(\mathcal{I}) \vert \cong \vI \rightarrow \vO$. 
	
	To show that it is a chromatic function, we start with the color preservation property. Notice that, since $\mu$ is simplicial and chromatic, $\chi(\sigma) = \chi(\mu (\sigma))$ for any simplex $\sigma \in \I$. This implies that, for any $x \in \vI, \vchi(x) = \vchi(\vmu(x))$. Moreover, it follows that, for any point $x$ and any simplicial neighborhood $N$ of $x$ of dimension $r$, $\vchi(N) = \vchi(\vmu(N))$. Since $\sigma$ must be of dimension at least $r$ since $N$ has dimension $r$, it follows that $\bigl|\vchi(N)\bigr| = \bigl|\vchi(N) \cap \vchi(\vmu (N))\bigr| \geq r+1$.
	
\end{proof}

We further develop the discrete/continuous duality by defining continuous tasks. Since chromatic functions correspond to chromatic simplicial maps, and chromatic simplicial maps determine task solvability, we can also use chromatic functions to express solvable tasks. 

\begin{definition}[Continuous task] \label{def:contas}
We say that a triple $\langle   \mathcal{I} ,  \mathcal{O}  , f \rangle $ is a continuous task if $\mathcal{I}$ and $\mathcal{O}$ are pure chromatic simplicial complexes of the same dimension, and $f : \vert \mathcal{I}\vert \rightarrow \vert \mathcal{O} \vert$ is a chromatic function.
\end{definition}

In order for our task definition to be complete, we also need to define a criterion for task solvability. A continuous task is solvable in ASM if there exists an  algorithm $\mathcal{A}$ with a decision map $\mu$ that ``approximates'' the chromatic function of the continuous task. Therefore we first need a formal definition for a chromatic approximation.

\begin{definition}[Chromatic approximation] \label{def:chrapp}
Let $f: \vert \mathcal{I} \vert \rightarrow \vert \mathcal{O} \vert$ be a chromatic function. We say that a chromatic simplicial map $\mu: \textrm{Sub}(\mathcal{I}) \rightarrow \mathcal{O}$ is a chromatic approximation to $f$ if, for all $\sigma \in \textrm{Sub}(\mathcal{I})$, 
$\mu (\sigma) \subseteq \carr(f(\vert \sigma \vert), \mathcal{O})$.
\end{definition}

Notice that while a chromatic approximation of $f$ is a more relaxed definition than a simplicial approximation of $f$ in the topological sense, as $f(\textrm{int}(\vsigma)) \subseteq \ostar(\mu(\sigma))$ need not hold, it adds a color preservation constraint.

\begin{definition}[ASM Continuous Task Solvability] \label{def:tassol}
We say that an algorithm $\mathcal{A}$ in ASM solves a continuous task $T = \langle \mathcal{I}, \mathcal{O}, f \rangle$, if $\mathcal{A}$ induces a subdivision $\textrm{Sub}(\mathcal{I})$ of $\mathcal{I}$, and a (simplicial) decision map $\mu: \textrm{Sub}(\mathcal{I}) \rightarrow \mathcal{O}$ that is a chromatic approximation, i.e., for each $\sigma \in \textrm{Sub} (\mathcal{I}), 
\mu (\sigma) \subseteq  \carr(f(\vert \sigma \vert), \mathcal{O})$.
\end{definition}

Another way to formulate this condition is by defining an induced task $\Delta_f$ associated to $f$. 
 
\begin{definition}[Induced Task] \label{def:indtask}
Given a continuous task $T = \langle \mathcal{I}, \mathcal{O}, f \rangle$, we define the task induced by $T$ as $T_f = \langle \mathcal{I}, \mathcal{O}, \Delta_f \rangle$, where $\Delta_f: \mathcal{I} \rightarrow 2^{\mathcal{O}}$ is the carrier map induced by $f$ as given by $\Delta_f(\sigma) = \{ \textrm{carr}(f(x),\mathcal{O}) \; \vert \; x \in \vert\sigma\vert\} = \textrm{carr}(f(|\sigma|),\mathcal{O})$.
\end{definition}

Recall from the ACT \cref{th:ACTif} that $T_f$ is solvable in ASM if there exists a subdivision $\Sub(\I)$ and a decision map $\mu:\Sub(\I) \to \O$ carried by $\Delta_f$. According \cref{def:tassol}, this indeed implies continuous task solvability for $T$ as well.


\subsection{The Chromatic Approximation Theorem}
\label{sec:chAppTh}

In the previous subsection, we provided the motivation and definitions for chromatic functions and continuous tasks as part of a discrete/continuous duality for ASM. We showed that chromatic simplicial maps generate chromatic functions. However, in order for this correspondence to be complete, we need to show that we can approximate any chromatic function with a simplicial chromatic map. In general algebraic topology, the simplicial approximation theorem allows us to discretize continuous functions. In the context of distributed computing, however,  we cannot apply the simplicial approximation theorem directly, since it does not necessarily preserve the color structure.

Therefore, in this subsection, we will prove that any chromatic function from a geometric pure simplicial complex $\vert \mathcal{I} \vert$ into a geometric pure simplicial complex $\vert \mathcal{O} \vert$ of the same dimension admits a chromatic approximation. To this end, we introduce the notion of the chromatic
projection for some color $c$. It maps interior points of a geometric simplex $\vert \sigma \vert$ to its border, by taking the ray $r$ from the vertex $v_i\in V(\sigma)$ with color $c$ to $x$, and mapping $x$ to the intersection of $r$ and the opposite border of $\vert \sigma \vert$. 

\begin{definition}[Chromatic projection]\label{def:chrproj}
Let $\mathcal{I}$ be a pure chromatic simplicial complex, and $c \in \chi(\mathcal{I})$. We define the chromatic projection with respect to $c$ as $\pi_c : \vI \setminus \vchi^{-1}(\{c\}) \rightarrow \vert \I \setminus \chi^{-1}(\{c\}) \vert$ as follows:
For $x\in \vI\setminus \vchi^{-1}(\{c\})$, let $\sigma = \carr(x,\I)=\{v_1, \ldots, v_n\}$ be the carrier of $x$ in $|\mathcal{I}|$. If $c \notin \chi(\sigma)$, we define $\pi_c(x) = x$. Now, assume that $ \chi(v_i) = c$, in which case we must have $n\geq 2$. Writing $x = \sum \limits_{k=1}^{n} \alpha_k \cdot v_k$, where each $v_k \in \sigma$, and the $\alpha_k$'s correspond to the affine coordinates of $x$ with respect to $\sigma$, we define 
$\displaystyle
\pi_c(x)= \sum \limits_{k \neq i} \frac{\alpha_k}{1- \alpha_i}  \cdot v_k$. 
\end{definition}

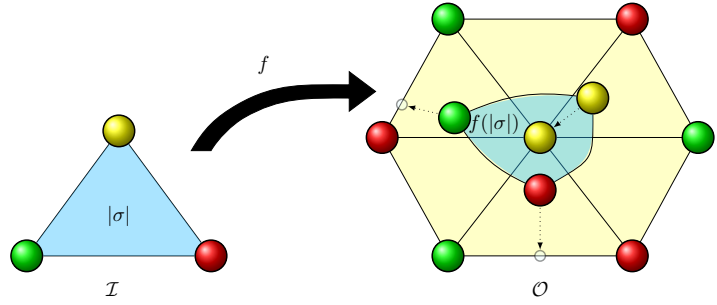
\begin{figure}[!t]
\centering
\scalebox{0.7}{	\begin{tikzpicture}
	\begin{pgfonlayer}{nodelayer}
		\node [style=VY] (9) at (10.5, 4.5) {};
		\node [style=VG] (0) at (-9, 0) {};
		\node [style=VR] (1) at (-2, 0) {};
		\node [style=VY] (2) at (-5.5, 4.75) {};
		\node [style=none] (4) at (-2, 0) {};
		\node [style=none] (5) at (-5.5, 4.75) {};
		\node [style=none] (6) at (-9, 0) {};
		\node [style=VG] (7) at (7, 0) {};
		\node [style=VR] (8) at (14, 0) {};
		\node [style=VR] (10) at (4.5, 4.5) {};
		\node [style=VG] (11) at (16.5, 4.5) {};
		\node [style=VR] (12) at (14, 9) {};
		\node [style=VG] (13) at (7, 9) {};
		\node [style=none] (14) at (7, 0) {};
		\node [style=none] (15) at (14, 0) {};
		\node [style=none] (16) at (16.5, 4.5) {};
		\node [style=none] (17) at (14, 9) {};
		\node [style=none] (18) at (7, 9) {};
		\node [style=none] (19) at (4.5, 4.5) {};
		\node [style=none] (21) at (-5.5, 1.5) {$|\sigma|$};
		\node [style=VG] (22) at (7.25, 5.25) {};
		\node [style=VR] (23) at (10.5, 2.5) {};
		\node [style=VY] (24) at (12.5, 6) {};
		\node [style=none] (26) at (10.5, 2.5) {};
		\node [style=none] (27) at (12.5, 6) {};
		\node [style=none] (28) at (7.25, 5.25) {};
		\node [style=none] (29) at (8.75, 5) {$f(|\sigma|)$};
		\node [style=none] (30) at (-2.75, 4) {};
		\node [style=none] (31) at (-1, 5.75) {};
		\node [style=none] (32) at (3.25, 6.5) {};
		\node [style=none] (33) at (-2, 4) {};
		\node [style=none] (34) at (-0.75, 5.25) {};
		\node [style=none] (35) at (3.25, 6) {};
		\node [style=none] (36) at (3.25, 7) {};
		\node [style=none] (37) at (3.25, 5.5) {};
		\node [style=none] (38) at (4.25, 6.25) {};
		\node [style=none] (39) at (0, 7.25) {$f$};
		\node [style=none] (40) at (-5.75, -1.25) {$\mathcal{I}$};
		\node [style=VBfaded] (42) at (5.25, 5.75) {};
		\node [style=VBfaded] (43) at (10.5, 0) {};
		\node [style=none] (45) at (10.5, -1.25) {$\mathcal{O}$};
	\end{pgfonlayer}
	\begin{pgfonlayer}{edgelayer}
		\draw (0) to (1);
		\draw (1) to (2);
		\draw (2) to (0);
		\draw [style=FaceBlue] (6.center)
			 to (4.center)
			 to (5.center)
			 to cycle;
		\draw [style=FaceYellow] (17.center)
			 to (18.center)
			 to (19.center)
			 to (14.center)
			 to (15.center)
			 to [in=240, out=60] (16.center)
			 to cycle;
		\draw (14.center) to (9);
		\draw (14.center) to (15.center);
		\draw (15.center) to (9);
		\draw (9) to (16.center);
		\draw [in=60, out=-120] (16.center) to (15.center);
		\draw (16.center) to (17.center);
		\draw (17.center) to (9);
		\draw (9) to (18.center);
		\draw (18.center) to (17.center);
		\draw (18.center) to (19.center);
		\draw (19.center) to (14.center);
		\draw (9) to (19.center);
		\draw [bend right=15] (22) to (23);
		\draw [bend right, looseness=1.50] (23) to (24);
		\draw [bend left=345] (24) to (22);
		\draw [style=FaceBlue] (28.center)
			 to [bend right=15] (26.center)
			 to [bend right, looseness=1.50] (27.center)
			 to [bend left=345] cycle;
		\draw [style=ArrowBlack] (38.center)
			 to (36.center)
			 to [in=90, out=-90] (32.center)
			 to [in=30, out=180] (31.center)
			 to [bend right=15] (30.center)
			 to (33.center)
			 to [bend left=15, looseness=0.75] (34.center)
			 to [in=180, out=30] (35.center)
			 to [in=90, out=-90] (37.center)
			 to cycle;
		\draw [style=Eoutd, in=345, out=165] (28.center) to (42);
		\draw [style=Eoutd] (26.center) to (43);
		\draw [style=Eoutd] (27.center) to (9);
	\end{pgfonlayer}
\end{tikzpicture}}
	\caption{The interior of a simplex is mapped to an open star of a vertex. All yellow vertices are mapped to the center of the star.}\label{fig:retract}
\end{figure}

An example for a chromatic projection can be found in \cref{fig:retract}, for $c = \textrm{yellow}$: Both the points $x$ marked by the green and the red inner node on the boundary of $f(|\sigma|)$ are mapped to the respective border of their carriers that lies opposite of the yellow vertex.

\cref{fig:retract} also illustrates the pivotal concept of star-covering introduced in \cref{def:starcov}, which requires the image of the interior of a simplex $\vsigma$ to be contained 
in the open star $\ostar(w)$ of some vertex $w\in V(\O)$.

\begin{definition}[Star-covered subdivision]\label{def:starcov}
Let $f : \vert \mathcal{I} \vert \rightarrow \vert \mathcal{O} \vert$ be a chromatic function. We say that a chromatic subdivision $\textrm{Sub}(\mathcal{I})$ is star-covered with respect to $f$ if for any $\sigma \in \Sub(\mathcal{I})$, $f(\interior(\vsigma))  \subseteq  \ostar(w)$ for some $w \in V(\mathcal{O})$.
\end{definition}

The following \cref{lemma:starcover} shows that a sufficiently deep chromatic subdivision
guarantees that a chromatic function $f$ will be star-covered with respect to $f$.

\begin{lemmarep}\label{lemma:starcover}
Let $f : \vert \mathcal{I} \vert \rightarrow \vert \mathcal{O} \vert$ be a chromatic function. Then, there exists $k \in \mathbb{N}$ such that $\Ch^k(\mathcal{I})$ is star-covered with respect to $f$.
\end{lemmarep}

\begin{proof}
Since $\mathcal{O}$ is pure and of the same dimension as $\mathcal{I}$, the collection $\mathcal{C} = \{  \ostar(w) \; \vert \; w \in \mathcal{O}\}$ is an open covering of $\vert \mathcal{O} \vert$. We claim that $\mathcal{C}^{-1} = \{ f^{-1}(U \cap f(\vI)) \; | \; U \in C \}$, the collection of preimages of $C$ under $f$ is a finite open cover for $\vert \mathcal{I} \vert $. Indeed, for every $x \in \vI$,  $f(x) \in \vO$. Since $\mathcal{C}$ is an open cover for $\vO$, there exists $U \in \mathcal{C}$ such that $f(x) \in U$. Since $U$ is open
and $f$ is continuous, the preimage of a sufficiently small neighborhood $N(f(x)) \subseteq (U \cap
f(\vI))$ is open and hence contained in $\mathcal{C}^{-1}$.

Since $\mathcal{I}$ is a finite simplicial complex, $\vI$ is a compact metric space. Therefore, there exists a Lebesgue number $\epsilon >0$ such that any set $S\subseteq \vI$ with a diameter less than $\epsilon$ is contained in an element of $\mathcal{C}^{-1}$. Since the chromatic subdivision is a mesh shrinking operation on $\vI$, there exists some $k \in \mathbb{N}$ such that any $\vsigma$ in $\vert \Ch^k(\mathcal{I})\vert$ has a diameter less than $\epsilon$. Therefore, for any $\sigma  \in \Ch^k(\mathcal{I})$, $\vsigma \in S$ for some $S \in \mathcal{C}^{-1}$. Consequently, $f(\vsigma) \subseteq \ostar(w)$ for some $w \in V(\mathcal{O})$, which confirms that $\Ch^k(\mathcal{I})$ is indeed star-covered with respect to $f$.
\end{proof}

A star-covered subdivision $\Sub(\mathcal{I})$ induces a coloring for the facets $\sigma \in \Sub(\mathcal{I})$. \cref{def:starcoloring} simply associates a facet $\sigma$ with the colors of all star centers that cover $f(\sigma)$ in $\O$. For example, in \cref{fig:retract}, $\sigma$ is assigned the yellow color here. This color assignment will be fundamental to the proof of \cref{thm:chrappr}.

\begin{definition}[Star coloring]\label{def:starcoloring}
Let $\Sub(\mathcal{I})$ be a star-covered subdivision with respect to $f: \vI \rightarrow \vO$. For $\sigma\in\mathcal{I}$, we define $\chi^*(\sigma) = \{ c \; \vert \; w \in V(\mathcal{O}), \; \chi(w) = c, \; f(\interior(\vsigma))  \subseteq  \ostar(w) \}$.
\end{definition}

\begin{definition}
Let $\Sub(\mathcal{I})$ be a star-covered subdivision with respect to $f: \vI \rightarrow \vO$ and $c \in \chi(\mathcal{O})$. We say that a facet $\sigma \in \Sub(\mathcal{I})$ is $c$-covered if $c \in \chi^*(\sigma)$, and define the $c$-subcomplex of $\Sub(\mathcal{I})$ as $\Sub(\mathcal{I})_c = \{\sigma \in  \Sub(\mathcal{I}) \; \vert \; c \in \chi^*(\sigma)\}$.
\end{definition}

The following quite obvious \cref{lemma:subref} shows that independently subdivided $c$-subcomplexes $\Sub(\mathcal{I})$ can be globally refined.

\begin{lemma}\label{lemma:subref}
Let $\mathcal{I}_1, \ldots, \mathcal{I}_k$ be subcomplexes of $\mathcal{I}$ such that $\bigcup\limits_{j=1}^k \mathcal{I}_j = \mathcal{I}$, and $\Sub(\mathcal{I}_j)$ be a chromatic subdivision of each $\mathcal{I}_j$. There exists a chromatic subdivision $\Sub(\mathcal{I})$ that refines each $\textrm{Sub}(\mathcal{I}_j)$.
\end{lemma}

Since a given facet $\sigma \in \Sub(\I)$ may be $c$-covered for several different colors $c$,
we pick one of those to obtain a color partition of $\Sub(\I)$, as provided by \cref{lemma:colpart}.

\begin{lemmarep}\label{lemma:colpart}
Let $\Sub(\mathcal{I})$ be a star-covered subdivision with respect to $f: \vI \rightarrow \vO$, 
with its induced $i$-colored subcomplexes $\Sub(\I)_i$, $1\leq i \leq m$. There exists a partition $\mathcal{P}=\{ P_1, \ldots, P_r \}$, $1\leq r \leq m$, of $\Sub(\mathcal{I})$ such that each $P_i \subseteq \Sub(\mathcal{I})_i$, and for any pair $i<j$, $P_j \cap \Sub(\mathcal{I})_i = \varnothing$. We call $\mathcal{P}$ the color partition of $\Sub(\mathcal{I})$.
\end{lemmarep}

\begin{proof}
Let $C_1 = \Sub(\mathcal{I})$, which is a star-covered subdivision with respect to $f$. According to \cref{lemma:starcover}, not all $i$-subcomplexes of $C_1$ are empty. As we can find a permutation of the coloring $\chi(\mathcal{I})$ that ensures $\Sub(\mathcal{I})_1 \neq \varnothing$, we can just define $P_1 = \Sub(\mathcal{I})_1$. 
Now consider $C_2 = \Sub(\mathcal{I}) \setminus P_1$. If $C_2 = \varnothing$, then $\mathcal{P} = \{ P_1\}$. Otherwise, we can proceed inductively to define the remaining $P_i$, $i\geq 2$.
\end{proof}

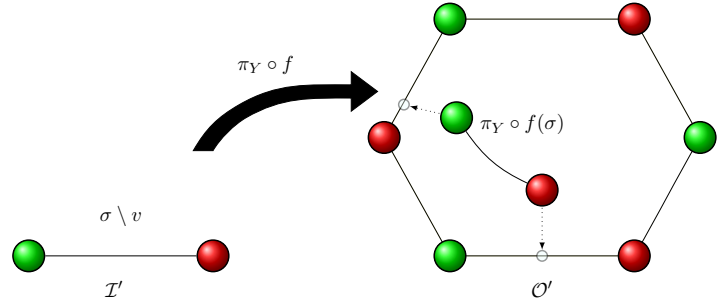
\begin{figure}[!t]
\centering
\scalebox{0.7}{	\begin{tikzpicture}
	\begin{pgfonlayer}{nodelayer}
		\node [style=VG] (0) at (-9, 0) {};
		\node [style=VR] (1) at (-2, 0) {};
		\node [style=none] (4) at (-2, 0) {};
		\node [style=none] (6) at (-9, 0) {};
		\node [style=VG] (7) at (7, 0) {};
		\node [style=VR] (8) at (14, 0) {};
		\node [style=VR] (10) at (4.5, 4.5) {};
		\node [style=VG] (11) at (16.5, 4.5) {};
		\node [style=VR] (12) at (14, 9) {};
		\node [style=VG] (13) at (7, 9) {};
		\node [style=none] (14) at (7, 0) {};
		\node [style=none] (15) at (14, 0) {};
		\node [style=none] (16) at (16.5, 4.5) {};
		\node [style=none] (17) at (14, 9) {};
		\node [style=none] (18) at (7, 9) {};
		\node [style=none] (19) at (4.5, 4.5) {};
		\node [style=none] (21) at (-5.5, 1.5) {$\sigma \setminus v$};
		\node [style=VG] (22) at (7.25, 5.25) {};
		\node [style=VR] (23) at (10.5, 2.5) {};
		\node [style=none] (26) at (10.5, 2.5) {};
		\node [style=none] (28) at (7.25, 5.25) {};
		\node [style=none] (29) at (9.75, 5) {$\pi_Y \circ f(\sigma)$};
		\node [style=none] (30) at (-2.75, 4) {};
		\node [style=none] (31) at (-1, 5.75) {};
		\node [style=none] (32) at (3.25, 6.5) {};
		\node [style=none] (33) at (-2, 4) {};
		\node [style=none] (34) at (-0.75, 5.25) {};
		\node [style=none] (35) at (3.25, 6) {};
		\node [style=none] (36) at (3.25, 7) {};
		\node [style=none] (37) at (3.25, 5.5) {};
		\node [style=none] (38) at (4.25, 6.25) {};
		\node [style=none] (39) at (0, 7.25) {$\pi_Y \circ f$};
		\node [style=none] (40) at (-5.75, -1.25) {$\mathcal{I}'$};
		\node [style=VBfaded] (42) at (5.25, 5.75) {};
		\node [style=VBfaded] (43) at (10.5, 0) {};
		\node [style=none] (45) at (10.5, -1.25) {$\mathcal{O}'$};
	\end{pgfonlayer}
	\begin{pgfonlayer}{edgelayer}
		\draw (0) to (1);
		\draw [style=FaceYellow] (17.center) to (18.center);
		\draw [style=FaceYellow] (18.center) to (19.center);
		\draw [style=FaceYellow] (19.center) to (14.center);
		\draw [style=FaceYellow] (14.center) to (15.center);
		\draw [style=FaceYellow, in=240, out=60] (15.center) to (16.center);
		\draw [style=FaceYellow] (16.center) to (17.center);
		\draw (14.center) to (15.center);
		\draw [in=60, out=-120] (16.center) to (15.center);
		\draw (16.center) to (17.center);
		\draw [in=180, out=0] (18.center) to (17.center);
		\draw (18.center) to (19.center);
		\draw (19.center) to (14.center);
		\draw [bend right=15] (22) to (23);
		\draw [style=ArrowBlack] (38.center)
			 to (36.center)
			 to [in=90, out=-90] (32.center)
			 to [in=30, out=180] (31.center)
			 to [bend right=15] (30.center)
			 to (33.center)
			 to [bend left=15, looseness=0.75] (34.center)
			 to [in=180, out=30] (35.center)
			 to [in=90, out=-90] (37.center)
			 to cycle;
		\draw [style=Eoutd, in=345, out=165] (28.center) to (42);
		\draw [style=Eoutd] (26.center) to (43);
	\end{pgfonlayer}
\end{tikzpicture}}
	\caption{The chromatic projection $\pi_Y$, applied to the chromatic function shown in \cref{fig:retract} with $v$ denoting the yellow (Y) vertex, induces a chromatic function in a lower dimension. The function $f(x)$ maps the green vertex $x \in \sigma\setminus{\{v\}}$, the red-green border of $\sigma$, to $f(x)$ represented by the green inner vertex, the carrier of which contains the yellow central vertex in \cref{fig:retract}. Applying $\pi_c(f(x)$ retracts this point to the opposite red-green border.} \label{fig:chromproj}
\end{figure}

The following \cref{lemma:chrprj} is instrumental in the proof of our chromatic simplicial approximation \cref{thm:chrappr}. It shows that chromatic projections applied to chromatic
functions provide chromatic functions, as illustrated in \cref{fig:chromproj}. 

\begin{lemmarep}\label{lemma:chrprj}
Let $f: \vI \rightarrow \vO$ be a chromatic function, and $\Sub(\mathcal{I})$ a star-covered subdivision with respect to $f$. Assume that $S_c$ is a $c$-colored subcomplex of $\Sub(\mathcal{I})$, and let $S_c' = S_c \setminus \{ v \in V(S_c) \; \vert \; \chi(v) = c \}$, $\mathcal{O}_c' = \mathcal{O} \setminus \{ v \in V(\mathcal{O}) \; \vert \; \chi(v) = c\}$. Then, $f_{c} : \vert S_c' \vert \rightarrow \vert \mathcal{O}_c' \vert$ defined by $f_{c}(x) = \pi_c(f (x))$ is a chromatic
function.
\end{lemmarep}

\begin{proof}
Let $V_c = \{ v \in V(\Sub(\mathcal{I})) \; | \; \chi(v)= c\}$. Notice that from \cref{def:chrproj}, $\pi_c$ is continuous in $\vert S_c  \setminus V_c \vert$. Since $f$ is continuous, $\pi_c \circ f$ must also be continuous in $\vert S_c' \vert \subset \vert S_c  \setminus V_c\vert$. Hence, 
$f_{c}(x) = \pi_c(f (x))$ is indeed continuous.

In order to show that $f_{c}(x)$ is chromatic, let $N$ be an $r$-dimensional simplicial neighborhood in $|S_c'|$. Note that $N$ is also an $r$-dimensional neighborhood in $\Sub(\mathcal{I})$. \cref{cor:subchr} implies that $f$ is also chromatic in $|S_c'|$, hence $|\vchi(N) \cap \vchi(f(N))|\geq r+1$ holds. Since $c \notin \vchi(N)$ 
as $N \subseteq |S_c'|$, it follows that $c \notin \vchi(N) \cap \vchi(f(N))$. Consequently, $  \vchi(N) \cap \vchi(f(N))  = ( \vchi(N) \cap \vchi(f(N)) ) \setminus \{c\} = \vchi(N) \setminus \{c\} \cap \vchi(f(N)) \setminus \{c\}$ by distributivity of set exclusion over intersection. Recalling $c \notin \vchi(N)$ and noticing that $\vchi (\pi_c \circ f (N)) = \vchi (f(N)) \setminus \{c\}$, it follows that $ \vchi(N) \cap \vchi(f(N)) = \vchi (N) \setminus \{c\} \cap \vchi(f(N)) \setminus \{c\} = \vchi (N)  \cap \vchi (\pi_c \circ f (N))$. We conclude that $|\vchi (N)  \cap \vchi (f_{c}(N))| = |\vchi(N) \cap \vchi(f(N))|\geq r+1$. Therefore, $f_{c}$ is indeed a chromatic function.
\end{proof}

The following \cref{lemma:grachrapp} is just the chromatic approximation theorem \cref{thm:chrappr} written out for 1-dimensional simplicial complexes. It will serve as the induction basis
for the proof of the latter.

\begin{lemmarep}\label{lemma:grachrapp}
Let $f :\vI \rightarrow \vO$ be a continuous chromatic function such that $\mathcal{I}$ and $\mathcal{O}$ are pure simplicial complexes of dimension $1$. There exists a chromatic subdivision $\Sub(\mathcal{I})$ of $\mathcal{I}$, and a chromatic simplicial map $\mu :\Sub(\mathcal{I} )\rightarrow \mathcal{O}$ that is a chromatic approximation of $f$.
\end{lemmarep}

\begin{proof}
From  \cref{lemma:starcover}, it follows that there exists a chromatic subdivision $\Sub(\I)$ which is star-covered with respect to $f$. We define a simplicial vertex map $\mu : \Sub (\I)\rightarrow \O$ as follows: For a vertex $v$, we set $\mu(v) = w $ if $\chi(w) = \chi(v)$ and $w \in \carr(f(v),\O)$. That is, $\mu$ maps vertex $v$ to the vertex of the same color in the carrier of $f(v)$, so obviously is a chromatic map. 

Now consider a facet $\sigma = \{v_1, v_2\} \in \Sub (\I)$. Since $f$ is star-covered, $f(\interior(\vsigma)) \subseteq \ostar(w_1)$ for some $w_1 \in \O$. Assume w.l.o.g.\ that $\mu(v_1) = w_1$, and let $w_2 = \mu(v_2)$. Notice that $w_2 \in \cstar(w_1)$, as otherwise $f(\interior(\vsigma)) \subseteq \ostar(w_1)$ is impossible. Therefore, there exists a simplex $\tau \in \O$ that includes $w_2$ and $w_1$. It follows that $\tau = \mu(\sigma) \in \O$, which proves that $\mu$ is a chromatic simplicial map. It follows from the definition of $\mu$ that it is a chromatic approximation to $f$.
\end{proof}

\begin{theoremrep}[Chromatic Approximation Theorem] \label{thm:chrappr}
Let $f :\vI \rightarrow \vO$ be a chromatic function such that $\mathcal{I}$ and $\mathcal{O}$ are pure simplicial complexes of dimension $k$, and with the same color set $\chi(\mathcal{I}) = \chi (\mathcal{O})$. There exists a chromatic subdivision $\Sub(\mathcal{I})$ of $\mathcal{I}$, and a chromatic simplicial map $\mu :\Sub(\mathcal{I} )\rightarrow \mathcal{O}$ that is a chromatic approximation of $f$.
\end{theoremrep}

\begin{proof}
We induction over $\textrm{dim}(\mathcal{I}) = \textrm{dim}(\mathcal{O})= k$.
The induction base $k = 1$ is provided by \cref{lemma:grachrapp}.
For our induction hypothesis, we assume that \cref{thm:chrappr} holds for arbitrary chromatic functions with respect to $\I'$ and $\O'$ with dimension less than $k\geq 2$. Let $f: \vI \rightarrow \vO$ be a chromatic function with respect to $\mathcal{I}$ and $\mathcal{O}$, where $\dim(\mathcal{I}) = \dim(\mathcal{O})= k$ and $\chi(\mathcal{I}) = \chi (\mathcal{O}) = \{1, \ldots,k+1\}$.

It follows from \cref{lemma:starcover} that there exists a chromatic subdivision $\textrm{Sub}^0(\mathcal{I})$ of $\mathcal{I}$, which is star-covered with respect to $f$. \cref{lemma:colpart}
thus ensures that there is a color partition $\mathcal{P}^0 = \{ \mathcal{K}^0_1, \ldots, \mathcal{K}^0_r \}$ of $\Sub^0(\mathcal{I})$. Recall that each $\mathcal{K}^0_i$ is an $i$-covered subcomplex of $\Sub^0(\mathcal{I})$. Therefore, for any $\sigma \in \mathcal{K}^0_i$, $f(\interior(\vsigma)) \subseteq \ostar(w)$ for some $w \in \mathcal{O}$ with color $i$. Now, iteratively for $i=1,2,\dots,r$, we will perform the following two steps:

As our first step for $i$, we define a partial vertex map $\mu^0_i: \mathcal{K}^0_i \to \O$ for vertices with color $i$, as illustrated for the small yellow vertex on the boundary of $f(\vsigma)$ in \cref{fig:stepone}: Let $v \in V(\mathcal{K}^0_i)$ be such that $\chi(v) = i$. Since $\mathcal{K}^0_i$ is $i$-covered, there exists some $w \in V(\mathcal{O})$ such that $\chi(w) = i$ and $f(v) \in \ostar(w)$. We choose any such $w$, and set $\mu^0_i(v)=w$. 
Note carefully that this partial map $\mu^0_i$ defines uniquely a partial map for vertices of color $i$ also in further chromatic subdivisions of $\mathcal{K}^0_i$: Let $\mathcal{L}_i$ be any such subdivision of $\mathcal{K}^0_i$ and $v \in V(\mathcal{L}_i)$, $\chi(v)=i$. Taking $v$ as a point in the corresponding 0-dimensional geometric simplex in $\mathcal{L}_i$, consider $\sigma = \carr(v,\mathcal{K}^0_i)$. Since $f$ is chromatic also in $\vert \mathcal{L}_i \vert$ by \cref{cor:subchr}, there exists a unique vertex $v' \in \sigma$ such that $\chi (v') = i$. We can therefore consistently define $\mu^0_i(v) = \mu^0_i(v')$.

\begin{figure}[!t]
	\centering
\scalebox{0.7}{\begin{tikzpicture}
	\begin{pgfonlayer}{nodelayer}
		\node [style=VY] (9) at (10.5, 4.5) {};
		\node [style=VGsmall] (0) at (-9, 0) {};
		\node [style=VRsmall] (1) at (-2, 0) {};
		\node [style=VYsmall] (2) at (-5.5, 4.75) {};
		\node [style=none] (4) at (-2, 0) {};
		\node [style=none] (5) at (-5.5, 4.75) {};
		\node [style=none] (6) at (-9, 0) {};
		\node [style=VG] (7) at (7, 0) {};
		\node [style=VR] (8) at (14, 0) {};
		\node [style=VR] (10) at (4.5, 4.5) {};
		\node [style=VG] (11) at (16.5, 4.5) {};
		\node [style=VR] (12) at (14, 9) {};
		\node [style=VG] (13) at (7, 9) {};
		\node [style=none] (14) at (7, 0) {};
		\node [style=none] (15) at (14, 0) {};
		\node [style=none] (16) at (16.5, 4.5) {};
		\node [style=none] (17) at (14, 9) {};
		\node [style=none] (18) at (7, 9) {};
		\node [style=none] (19) at (4.5, 4.5) {};
		\node [style=VGsmall] (22) at (7.25, 5.25) {};
		\node [style=VRsmall] (23) at (10.5, 2.5) {};
		\node [style=VYsmall] (24) at (12.5, 6) {};
		\node [style=none] (26) at (10.5, 2.5) {};
		\node [style=none] (27) at (12.5, 6) {};
		\node [style=none] (28) at (7.25, 5.25) {};
		\node [style=none] (30) at (-2.75, 4) {};
		\node [style=none] (31) at (-1, 5.75) {};
		\node [style=none] (32) at (3.25, 6.5) {};
		\node [style=none] (33) at (-2, 4) {};
		\node [style=none] (34) at (-0.75, 5.25) {};
		\node [style=none] (35) at (3.25, 6) {};
		\node [style=none] (36) at (3.25, 7) {};
		\node [style=none] (37) at (3.25, 5.5) {};
		\node [style=none] (38) at (4.25, 6.25) {};
		\node [style=none] (39) at (0, 7.25) {$f$};
		\node [style=none] (40) at (-5.75, -1.25) {$\Sub^0(\mathcal{I})$};
		\node [style=VBfaded] (42) at (5.25, 5.75) {};
		\node [style=VBfaded] (43) at (10.5, 0) {};
		\node [style=none] (45) at (10.5, -1.25) {$\mathcal{O}$};
		\node [style=VRsmall] (46) at (-8, 0) {};
		\node [style=VGsmall] (47) at (-3, 0) {};
		\node [style=VRsmall] (48) at (7.5, 4.75) {};
		\node [style=VGsmall] (49) at (10, 2.75) {};
		\node [style=VBfaded] (50) at (4.75, 5) {};
		\node [style=VBfaded] (51) at (9, 0) {};
		\node [style=VGsmall] (52) at (8.25, 4) {};
		\node [style=VRsmall] (53) at (8.75, 3.5) {};
		\node [style=VBfaded] (54) at (5.25, 3.25) {};
		\node [style=VBfaded] (55) at (6, 1.75) {};
		\node [style=VGsmall] (56) at (-6.25, 0) {};
		\node [style=VRsmall] (57) at (-4.75, 0) {};
	\end{pgfonlayer}
	\begin{pgfonlayer}{edgelayer}
		\draw (0) to (1);
		\draw (1) to (2);
		\draw (2) to (0);
		\draw [style=FaceBlue] (6.center)
			 to (4.center)
			 to (5.center)
			 to cycle;
		\draw [style=FaceYellow] (17.center)
			 to (18.center)
			 to (19.center)
			 to (14.center)
			 to (15.center)
			 to [in=240, out=60] (16.center)
			 to cycle;
		\draw (14.center) to (9);
		\draw (14.center) to (15.center);
		\draw (15.center) to (9);
		\draw (9) to (16.center);
		\draw [in=60, out=-120] (16.center) to (15.center);
		\draw (16.center) to (17.center);
		\draw (17.center) to (9);
		\draw (9) to (18.center);
		\draw (18.center) to (17.center);
		\draw (18.center) to (19.center);
		\draw (19.center) to (14.center);
		\draw (9) to (19.center);
		\draw [bend right=15] (22) to (23);
		\draw [bend right, looseness=1.50] (23) to (24);
		\draw [bend left=345] (24) to (22);
		\draw [style=FaceBlue] (28.center)
			 to [bend right=15] (26.center)
			 to [bend right, looseness=1.50] (27.center)
			 to [bend left=345] cycle;
		\draw [style=ArrowBlack] (38.center)
			 to (36.center)
			 to [in=90, out=-90] (32.center)
			 to [in=30, out=180] (31.center)
			 to [bend right=15] (30.center)
			 to (33.center)
			 to [bend left=15, looseness=0.75] (34.center)
			 to [in=180, out=30] (35.center)
			 to [in=90, out=-90] (37.center)
			 to cycle;
		\draw [style=Eoutd, in=345, out=165] (28.center) to (42);
		\draw [style=Eoutd] (26.center) to (43);
		\draw [style=Eoutd] (27.center) to (9);
		\draw (46) to (5.center);
		\draw (47) to (5.center);
		\draw [bend left=15, looseness=1.25] (48) to (27.center);
		\draw [bend left=15, looseness=1.25] (27.center) to (49);
		\draw [style=Eoutd] (49) to (51);
		\draw [style=Eoutd] (48) to (50);
		\draw [style=Eoutd] (52) to (54);
		\draw [style=Eoutd] (53) to (55);
		\draw (56) to (5.center);
		\draw (57) to (5.center);
		\draw [bend left=15] (52) to (27.center);
		\draw [bend right=15] (53) to (27.center);
	\end{pgfonlayer}
\end{tikzpicture}}
	\caption{Star-cover for the chromatic subdivision $\Sub^0(\I)$, which induces a trivial color partition consisting of a single $\mathcal{K}^0_1$, with color~$1$ representing yellow, containing all images of the subdivides simplices in $\Sub^0(\I)$. The small yellow vertex is mapped by $\mu^0_1$ to the yellow center of the star, the small read and green vertices 
will be retracted to the opposite border and appropriately mapped in the induction step in the proof of \cref{thm:chrappr}, and finally mapped to some border vertices as shown in \cref{fig:chrapprx}.}\label{fig:stepone}
\end{figure}

\begin{figure}[h!]
\centering
\scalebox{0.7}{	\begin{tikzpicture}
	\begin{pgfonlayer}{nodelayer}
		\node [style=VY] (9) at (10.5, 4.5) {};
		\node [style=VGsmall] (0) at (-9, 0) {};
		\node [style=VRsmall] (1) at (-2, 0) {};
		\node [style=VYsmall] (2) at (-5.5, 4.75) {};
		\node [style=none] (4) at (-2, 0) {};
		\node [style=none] (5) at (-5.5, 4.75) {};
		\node [style=none] (6) at (-9, 0) {};
		\node [style=VG] (7) at (7, 0) {};
		\node [style=VR] (8) at (14, 0) {};
		\node [style=VR] (10) at (4.5, 4.5) {};
		\node [style=VG] (11) at (16.5, 4.5) {};
		\node [style=VR] (12) at (14, 9) {};
		\node [style=VG] (13) at (7, 9) {};
		\node [style=none] (14) at (7, 0) {};
		\node [style=none] (15) at (14, 0) {};
		\node [style=none] (16) at (16.5, 4.5) {};
		\node [style=none] (17) at (14, 9) {};
		\node [style=none] (18) at (7, 9) {};
		\node [style=none] (19) at (4.5, 4.5) {};
		\node [style=VGsmall] (22) at (8, 5.25) {};
		\node [style=VRsmall] (23) at (10.5, 2.5) {};
		\node [style=VYsmall] (24) at (12.5, 6) {};
		\node [style=none] (26) at (10.5, 2.5) {};
		\node [style=none] (27) at (12.5, 6) {};
		\node [style=none] (28) at (8, 5.25) {};
		\node [style=none] (30) at (-2.75, 4) {};
		\node [style=none] (31) at (-1, 5.75) {};
		\node [style=none] (32) at (3.25, 6.5) {};
		\node [style=none] (33) at (-2, 4) {};
		\node [style=none] (34) at (-0.75, 5.25) {};
		\node [style=none] (35) at (3.25, 6) {};
		\node [style=none] (36) at (3.25, 7) {};
		\node [style=none] (37) at (3.25, 5.5) {};
		\node [style=none] (38) at (4.25, 6.25) {};
		\node [style=none] (39) at (0, 7.25) {$f$};
		\node [style=none] (40) at (-5.75, -1.25) {$\Sub^0(\mathcal{I})$};
		\node [style=VBfaded] (42) at (5.25, 5.75) {};
		\node [style=VBfaded] (43) at (10.5, 0) {};
		\node [style=none] (45) at (10.5, -1.25) {$\mathcal{O}$};
		\node [style=VRsmall] (46) at (-8, 0) {};
		\node [style=VGsmall] (47) at (-3, 0) {};
		\node [style=VRsmall] (48) at (8.25, 4.75) {};
		\node [style=VGsmall] (49) at (10, 2.75) {};
		\node [style=VBfaded] (50) at (4.75, 5) {};
		\node [style=VBfaded] (51) at (9, 0) {};
		\node [style=none] (59) at (10.5, 4.5) {};
		\node [style=none] (60) at (7, 9) {};
		\node [style=none] (61) at (4.5, 4.5) {};
		\node [style=none] (62) at (10.5, 4.5) {};
		\node [style=none] (63) at (4.5, 4.5) {};
		\node [style=none] (64) at (7, 0) {};
		\node [style=none] (65) at (14, 0) {};
		\node [style=none] (66) at (10.5, 4.5) {};
		\node [style=none] (67) at (7, 0) {};
		\node [style=VGsmall] (68) at (8.75, 4) {};
		\node [style=VRsmall] (69) at (9.25, 3.5) {};
		\node [style=VBfaded] (70) at (5.25, 3.25) {};
		\node [style=VBfaded] (71) at (6, 1.75) {};
		\node [style=VGsmall] (72) at (-6.25, 0) {};
		\node [style=VRsmall] (73) at (-4.75, 0) {};
		\node [style=none] (74) at (-8.25, 0.5) {$\sigma_1$};
		\node [style=none] (75) at (-6.75, 0.5) {$\sigma_2$};
		\node [style=none] (76) at (-2.75, 0.5) {$\sigma_5$};
		\node [style=none] (77) at (-5.5, 0.5) {$\sigma_3$};
		\node [style=none] (78) at (-4, 0.5) {$\sigma_4$};
		\node [style=none] (79) at (6.75, 5.5) {$\mu^1(\sigma_1)$};
		\node [style=none] (80) at (6.75, 3.5) {$\mu^1(\sigma_2)$};
		\node [style=none] (81) at (10, 1.5) {$\mu^1(\sigma_5)$};
		\node [style=none] (82) at (6.75, 2.75) {$\mu^1(\sigma_3)$};
		\node [style=none] (83) at (6.75, 2) {$\mu^1(\sigma_4)$};
	\end{pgfonlayer}
	\begin{pgfonlayer}{edgelayer}
		\draw (0) to (1);
		\draw (1) to (2);
		\draw (2) to (0);
		\draw [style=FaceBlue] (6.center)
			 to (4.center)
			 to (5.center)
			 to cycle;
		\draw [style=FaceYellow] (17.center)
			 to (18.center)
			 to (19.center)
			 to (14.center)
			 to (15.center)
			 to [in=240, out=60] (16.center)
			 to cycle;
		\draw (14.center) to (9);
		\draw (14.center) to (15.center);
		\draw (15.center) to (9);
		\draw (9) to (16.center);
		\draw [in=60, out=-120] (16.center) to (15.center);
		\draw (16.center) to (17.center);
		\draw (17.center) to (9);
		\draw (9) to (18.center);
		\draw (18.center) to (17.center);
		\draw (18.center) to (19.center);
		\draw (19.center) to (14.center);
		\draw (9) to (19.center);
		\draw [bend right=15] (22) to (23);
		\draw [bend right, looseness=1.50] (23) to (24);
		\draw [bend left=345] (24) to (22);
		\draw [style=FacePur] (28.center)
			 to [bend right=15] (26.center)
			 to [bend right, looseness=1.50] (27.center)
			 to [bend left=345] cycle;
		\draw [style=ArrowBlack] (38.center)
			 to (36.center)
			 to [in=90, out=-90] (32.center)
			 to [in=30, out=180] (31.center)
			 to [bend right=15] (30.center)
			 to (33.center)
			 to [bend left=15, looseness=0.75] (34.center)
			 to [in=180, out=30] (35.center)
			 to [in=90, out=-90] (37.center)
			 to cycle;
		\draw [style=Eoutd] (27.center) to (9);
		\draw (46) to (5.center);
		\draw (47) to (5.center);
		\draw [bend left=15, looseness=1.25] (48) to (27.center);
		\draw [bend left=15, looseness=1.25] (27.center) to (49);
		\draw [style=FaceBlue] (61.center) to (59.center);
		\draw [style=FaceBlue] (59.center) to (60.center);
		\draw [style=FaceBlue] (60.center) to (61.center);
		\draw [style=FaceBlue] (61.center)
			 to (59.center)
			 to (60.center)
			 to cycle;
		\draw [style=FaceBlue] (64.center) to (62.center);
		\draw [style=FaceBlue] (62.center) to (63.center);
		\draw [style=FaceBlue] (63.center) to (64.center);
		\draw [style=FaceBlue] (64.center)
			 to (62.center)
			 to (63.center)
			 to cycle;
		\draw [style=FaceBlue] (67.center) to (65.center);
		\draw [style=FaceBlue] (65.center) to (66.center);
		\draw [style=FaceBlue] (66.center) to (67.center);
		\draw [style=FaceBlue] (67.center)
			 to (65.center)
			 to (66.center)
			 to cycle;
		\draw [style=Eoutd] (22) to (13);
		\draw [style=Eoutd] (48) to (10);
		\draw [style=Eoutd] (23) to (8);
		\draw [style=Eoutd] (49) to (7);
		\draw [style=Eoutd] (68) to (14.center);
		\draw [style=Eoutd] (69) to (19.center);
		\draw [bend left=15] (68) to (27.center);
		\draw [bend right=15] (69) to (27.center);
		\draw (72) to (5.center);
		\draw (73) to (5.center);
	\end{pgfonlayer}
\end{tikzpicture}}
	\caption{Mapping of the small red and green vertices in \cref{fig:stepone}, which were retracted to the outer border of the star, to appropriate border vertices via $\mu^1_2$ and $\mu^1_3$ (denoted as $\mu^1$ in the figure for brevity) in the induction step in the proof of \cref{thm:chrappr}.}
	\label{fig:chrapprx}
\end{figure}

As our second step for $i$, we define a partial vertex map $\mu^1_i: \mathcal{K}^0_i \to \O$ for vertices with a color $\neq i$, as illustrated for the small red and green vertices on the boundary of $f(\vsigma)$ in \cref{fig:stepone}. Let $\mathcal{K}^1_i = \mathcal{K}^0_i \setminus \{ v \in V(\mathcal{K}^0_i) \; \vert \; \chi(v) = i \}$ and $\mathcal{O}_i = \mathcal{O} \setminus \{ v \in V(\mathcal{O}) \; \vert \; \chi(v) = i \}$.
Since $f$ is chromatic, \cref{lemma:chrprj} implies that each $f_i : \vert \mathcal{K}^1_i \vert \rightarrow \vert \mathcal{O}_i \vert$ defined by $f_i(x) = \pi_i (f(x))$ is also chromatic. Notice that each $\mathcal{K}^1_i$ and $\mathcal{O}_i$ has dimension less than $k$. Therefore, the induction hypothesis holds, and there exist $\mu^1_i : \Sub(\mathcal{K}^1_i) \rightarrow \mathcal{O}_i$ that are chromatic approximations to $f_i$, as illustrated in \cref{fig:chrapprx}. As before, each $\mu^1_i$ defines a vertex map for any further chromatic subdivision $\mathcal{L}_i$ of $\Sub(\mathcal{K}^1_i)$ in the same way 
as $\mu^0_i$ did for further subdivisions of $\mathcal{K}^0_i$.

Since each $\mu^1_i$ defines a vertex map for vertices of color different to $i$, and $\mu^0_i$ defines a vertex map for vertices of color $i$, the join $\mu^1_i * \mu^0_i = \mu_i$ defines
a vertex map for $\mathcal{K}^0_i$ for all colors. 

\medskip

To combine all these partial functions $\mu_i$ into a global function $\mu$, we make use of \cref{lemma:subref}. It ensures that there is a chromatic subdivision $\mathcal{K}$ of $\Sub(\mathcal{I})$ that refines every $\Sub(\mathcal{K}^0_i)$. Recall that every $\mu_i$ is well defined for any further chromatic subdivision $\Sub(\mathcal{K}^0_i)$. Therefore, we can globally define $\mu: \mathcal{K} \rightarrow \mathcal{O}$ as $\mu(v) = \mu_i(v)$, where $i$ is the smallest color such that $v \in \Sub(\mathcal{K}^0_i)$ in a sufficiently deep chromatic subdivision of $\mathcal{K}^0_i$. 

In only remains to prove that the so constructed $\mu$ is a chromatic approximation to $f$. It suffices to show that $\mu(\sigma) \subseteq \carr(f(\vsigma), \mathcal{O})$ for every facet $\sigma = \{ v_1, \ldots, v_n \} \in \mathcal{K}$. Let $\carr(f(\vsigma), \mathcal{O})= \kappa = \{w_1, \ldots ,w_n \}$. Since it generally holds for facets in a pure simplicial complex that $\kappa = \bigcap \limits_{i=1}^n \cstar(w_i)$, we only need to prove that $\mu(\sigma) \subseteq \cstar(w_i)$ for each $w_i$. This follows from the inductive construction of each $\mu_i$ and $\mu$ and the properties of the elements of color partition $\mathcal{P}^0$, however, which are based
on star-covered subdivisions.
\end{proof}



\begin{toappendix}
\section{Combinatorial Topology Preliminaries}\label{sec:topprelims}

For ease of reference, we provide a collection of basic combinatorial topology definitions, which have primarily been taken from~\cite{HKR13}. \cref{sec:combtop} is dedicated to basic combinatorial topology, \cref{sec:distcomp} adds definitions required for the modeling of a distributed systems using combinatorial topology.

\subsection{Combinatorial Topology}
\label{sec:combtop}

The first definition that we need is the definition of an abstract simplicial complex, which can can be thought of as a ``high-dimensional'' graph. Abstract simplicial complexes have the advantage of having a discrete combinatorial construction, while at the same time, they are able to express triangulated manifolds.

\begin{definition}[Abstract simplicial complex]\label{def:abssimpcomplex}
	An abstract simplicial complex $\mathcal K$ is a pair $\langle V(\mathcal{K}), F(\mathcal{K}) \rangle$, where $V(\mathcal{K})$ is a set, $F(\mathcal{K}) \subseteq 2^{V(\mathcal{K})}$, and for any $\sigma, \tau \in 2^{V(\mathcal{K})}$ such that $\sigma \subseteq \tau$ and  $\tau \in F(\mathcal{K})$, then $\sigma \in F(\mathcal{K})$. $V(\mathcal{K})$ is called the vertex set, and $F(\mathcal{K})$ is the face set of $\mathcal{K}$. The elements of $V(\mathcal{K})$ are called the vertices, and the elements of $F(\mathcal{K})$ are called the faces or simplices. We say that an abstract simplicial complex is \emph{finite} if its vertex set is finite.
	
	We say that a simplex $\sigma$ is a \emph{facet} if it is maximal with respect to containment.
\end{definition}

Traditionally, dimension is a number that represents how many axes we need to describe a point in space. The dimension of a simplex of an abstract simplicial complex is just the number of its vertices minus 1.

\begin{definition}[Dimension] \label{def:dimension}
	Let $\mathcal{K}$ be an abstract simplicial complex, and $\sigma \in F(\mathcal{K})$ be a simplex. We say that $\sigma$ has dimension $k$, denoted by $\textrm{dim}(\sigma)=k$, if it has a cardinality of $k+1$. We say that $\mathcal{K}$ is of dimension $k$ if it has a simplex of maximum dimension $k$.

\end{definition}

In the context of distributed computing, local processes' states correspond to vertices, and global configurations correspond to faces. Consequently, we only consider simplicial complexes where all simplices are of the same dimension. We call this particular type of simplicial complexes \emph{pure abstract simplicial complexes}.

Now that we have defined the basic objects in combinatorial topology, we can define the morphisms that preserve the structure. 

\begin{definition}[Simplicial maps] \label{def:simpmap}
	Let $\mathcal{K}$ and $\mathcal{L}$ be abstract simplicial complexes. We say that a vertex map $\mu : V(\mathcal{K}) \rightarrow V(\mathcal{L})$ is a simplicial map if for any $\sigma \in F(\mathcal{K}), \mu (\sigma) \in \mathcal{L}$.
	
	We say that a simplicial map $\mu$ is \emph{rigid} if for any $\sigma$ of dimension $d$, $\mu(\sigma)$ is also of dimension $d$
\end{definition}

It should be noted that simplicial maps are the discrete analogon to continuous functions. This equivalence is formally stated through the simplicial approximation theorem.

In the context of distributed systems, it is sometimes inevitable to enrich simplicial complexes with a coloring. Such a coloring corresponds to extracting the processes' ids from the local states.

\begin{definition}[Coloring]\label{def:coloring}
	Let $\mathcal{K}$ be a finite abstract simplicial complex of dimension $k$. We say that a function $\chi: V(\mathcal{K}) \rightarrow \{1,2, \ldots, k+1\}$ is a proper coloring if for any simplex $\sigma \in F(\mathcal{K})$, $\chi$ is injective at $\sigma$. 
\end{definition}

In order to simplify notation, whenever we have two disjoint abstract simplicial complexes $\mathcal{K}$ and $\mathcal{L}$, both of dimension $k$, with colorings $\mu_1$ and $\mu_2$, we will instead implicitly consider a "global" coloring $\mu: V(\mathcal{K}) \cup V(\mathcal{L}) \rightarrow \{1, \ldots, k+1\}$, since it is simpler and usually free from ambiguity.

Since we are interested that morphisms preserve the color structure, we must also add a coloring restriction.

\begin{definition}[Chromatic simplicial maps]\label{def:chrsimpmap}
	Let $\mathcal{K}$ resp.\ $\mathcal{L}$ be abstract simplicial complexes of dimension $k$, and 
$\chi_1$ resp.\ $\chi_2$ proper colorings for $\mathcal{K}$ resp.\ $\mathcal{L}$. We say that a simplicial map is chromatic if for any $v \in V(\mathcal{K}),\; \chi_1(v)) = \chi_2(\mu(v))$. 
	
	It should be noted that all chromatic simplicial maps are rigid.
\end{definition}

We need to define a substructure relation. Notice that since an abstract simplicial complex is a pair of two sets, it is natural to define the substructure relation based on set containment.
\begin{definition}[Subcomplex]
	Let $\mathcal{K}$ and $\mathcal{L}$ be abstract simplicial complexes, we say that $\mathcal{L}$ is a subcomplex of $\mathcal{K}$ if $V(\mathcal{L}) \subseteq V(\mathcal{K})$ and $F(\mathcal{L}) \subseteq F(\mathcal{K})$.
\end{definition}

Since we have a definition for the subcomplex relation, we can now define the $k$-dimensional skeleton of an abstract simplicial complex. 

\begin{definition}[r-Skeleton] \label{def:skel}
	Let $\I$ be a $k$-dimensional abstract simplicial complex, and $r \leq k$. We define the $r$-skeleton of $I$, $\textrm{Skel-}r(\I)$ as the subcomplex induced by the $r$-dimensional simplices of $\I$. More precisely:
	\begin{align}
		V(\textrm{Skel-}r(\I)) &= \{v \in V(\I) \; | \; \exists \: \sigma \in F(\I) \wedge \textrm{dim}(\sigma) = r \wedge v \in \sigma \}\nonumber,\\
		F(\textrm{Skel-}r(\I)) &= \{\sigma \in F(\I) \; | \; \exists \: \tau \in F(\I) \wedge \textrm{dim}(\tau) = r \wedge \sigma \subseteq \tau \}.\nonumber
	\end{align}
\end{definition}

\begin{definition}[Carrier Map]\label{def:carrmap}
	Let $\mathcal{K}$ and $\mathcal{L}$ be abstract simplicial complexes and $\Phi: \mathcal{K} \rightarrow 2^{\mathcal{L}}$. We say that $\Phi$ is a carrier map if $\Phi(\sigma)$ is a subcomplex of $\mathcal{L}$ for any $\sigma \in \mathcal{K}$, and for any $\sigma_1, \sigma_2 \in \mathcal{L}$, $\Phi( \sigma_1 \cap \sigma_2) \subseteq \Phi(\sigma_1) \cap \Phi(\sigma_2)$.
	
	We say that a carrier map $\Phi: \mathcal{K} \rightarrow 2^\mathcal{L}$
	is \emph{rigid} if for every simplex $\sigma \in \mathcal{K}$ of dimension $d$, the subcomplex $\Phi(\sigma)$ is pure of dimension $d$. It is is \emph{strict} if for every simplices $\sigma, \tau \in \mathcal{K}$, $\Phi(\sigma \cap \tau) = \Phi(\sigma) \cap \Phi(\tau)$. Finally, it	carries a simplicial vertex map $\mu: \mathcal{K} \rightarrow \mathcal{L}$ if, for any $\sigma \in \mathcal{K}$, $\mu(\sigma) \in \Phi (\sigma)$.
\end{definition}

\begin{definition}[Star of a vertex]\label{def:absstar}
	Let $v \in V(\mathcal K)$. We define the star of $v$ as the subcomplex of $\mathcal{K}$ of all simplices that contain $v$.
\end{definition}

Subdivisions are used to create refined simplicial complexes out of a given simplicial complex.
In the barycentric subdivision, the original faces become vertices, and simplex chains become the faces of the new subdivision. 

\begin{definition}[Barycentric subdivision]\label{def:barsub}
	Let $\mathcal{K}$ be an abstract simplicial complex. We define the barycentric subdivision, denoted by $\textrm{Bary}(\mathcal{K})$ through its vertex set and faces as follows.
\begin{align}	
	V(\textrm{Bary}(\mathcal{K})) &= F(\mathcal{K})\nonumber,\\
	F(\textrm{Bary}(\mathcal{K})) &= \{ (\sigma_1, \sigma_2, \ldots , \sigma_k) \; | \; \sigma_i \in F(\mathcal{K}) \; \wedge \; \sigma_1 \subseteq \sigma_2 \subseteq \ldots \subseteq \sigma_k\}.\nonumber
\end{align}	
\end{definition}
	
Intuitively, the barycentric subdivision adds a vertex in the center of each simplex, and joins each of the original vertices to the new vertex. Since it does not preserve colors, the more complex chromatic subdivision is typically used instead.

\begin{definition}[Chromatic subdivision]\label{def:chrsub}
	Let $\mathcal{K}$ be a $k$-dimensional abstract simplicial complex with a proper coloring $\chi: (V(\mathcal{K})) \rightarrow \{1, \ldots, k+1\}$. We define the chromatic subdivision through its vertex set and faces as follows.
\begin{align}		
	V(\textrm{Ch}(\mathcal{K})) &= \{(i,\sigma) \; | \; \sigma \in F(\mathcal{K}) \;  \wedge \; i \in \chi(\sigma)\}, \nonumber\\
	F(\textrm{Ch}(\mathcal{K})) &= \{(i_1, \sigma_1), (i_2, \sigma_2), \ldots (i_r, \sigma_r) \; | \;\sigma_j \in F(\mathcal{K}) \; \wedge \; \sigma_1 \subseteq \ldots \subseteq \sigma_r \; \wedge \; i_m \neq i_s \textrm{ for all } m \neq s\}.\nonumber
\end{align}	
\end{definition}

Intuitively, the chromatic subdivision is similar to the barycentric subdivision, but instead of only inserting a single vertex inside a face, it inserts a smaller face of the same dimension. See \cref{fig:subdivs} for illustrations of both the barycentric and the chromatic subdivision.

\begin{figure}[h!] \center
	\begin{tikzpicture}
	\begin{pgfonlayer}{nodelayer}
		\node [style=VGsmall] (0) at (-16, 0) {};
		\node [style=VRsmall] (1) at (-9, 0) {};
		\node [style=VYsmall] (2) at (-12.5, 5) {};
		\node [style=none] (11) at (-16, 0) {};
		\node [style=none] (12) at (-12.5, 5) {};
		\node [style=none] (13) at (-9, 0) {};
		\node [style=none] (22) at (-12.75, -1.75) {$\I$};
		\node [style=VBsmall] (23) at (-5, 0) {};
		\node [style=VBsmall] (24) at (2, 0) {};
		\node [style=VYsmall] (25) at (-1.5, 5) {};
		\node [style=none] (26) at (-5, 0) {};
		\node [style=VBsmall] (27) at (-1.5, 5) {};
		\node [style=none] (28) at (2, 0) {};
		\node [style=VBsmall] (29) at (-1.5, 2) {};
		\node [style=VBsmall] (30) at (-1.5, 0) {};
		\node [style=VBsmall] (31) at (0, 2.75) {};
		\node [style=VBsmall] (32) at (-3.25, 2.75) {};
		\node [style=none] (33) at (-1.25, -1.75) {$\Bary(\I)$};
		\node [style=none] (34) at (-13.5, 0.75) {};
		\node [style=VGsmall] (35) at (6, 0) {};
		\node [style=VRsmall] (36) at (13, 0) {};
		\node [style=VYsmall] (37) at (9.5, 5) {};
		\node [style=none] (38) at (6, 0) {};
		\node [style=none] (39) at (9.5, 5) {};
		\node [style=none] (40) at (13, 0) {};
		\node [style=VGsmall] (42) at (10.5, 2.25) {};
		\node [style=VRsmall] (43) at (8.5, 2.25) {};
		\node [style=VYsmall] (44) at (9.5, 0.75) {};
		\node [style=VGsmall] (45) at (8.25, 3.25) {};
		\node [style=VYsmall] (46) at (7.25, 1.75) {};
		\node [style=VGsmall] (47) at (10.75, 0) {};
		\node [style=VRsmall] (48) at (8.5, 0) {};
		\node [style=VRsmall] (49) at (10.75, 3.25) {};
		\node [style=VYsmall] (50) at (11.75, 1.75) {};
		\node [style=none] (51) at (9.5, -1.75) {$\Ch(\I)$};
	\end{pgfonlayer}
	\begin{pgfonlayer}{edgelayer}
		\draw (0) to (2);
		\draw (2) to (1);
		\draw (1) to (0);
		\draw [style=FaceBlue] (12.center)
			 to [in=120, out=-45, looseness=0.00] (13.center)
			 to (11.center)
			 to cycle;
		\draw (23) to (25);
		\draw (25) to (24);
		\draw (24) to (23);
		\draw [style=FaceBlue] (27.center)
			 to [in=120, out=-45, looseness=0.00] (28.center)
			 to (26.center)
			 to cycle;
		\draw (30) to (29);
		\draw (26.center) to (29);
		\draw (29) to (32);
		\draw (27) to (29);
		\draw (29) to (31);
		\draw (29) to (28.center);
		\draw (35) to (37);
		\draw (37) to (36);
		\draw (36) to (35);
		\draw [style=FaceBlue] (39.center)
			 to [in=120, out=-45, looseness=0.00] (40.center)
			 to (38.center)
			 to cycle;
		\draw (42) to (44);
		\draw (44) to (43);
		\draw (43) to (42);
		\draw (45) to (43);
		\draw (43) to (46);
		\draw (44) to (48);
		\draw (44) to (47);
		\draw (49) to (42);
		\draw (42) to (50);
		\draw (39.center) to (43);
		\draw (39.center) to (42);
		\draw (38.center) to (43);
		\draw (38.center) to (44);
		\draw (42) to (40.center);
		\draw (44) to (40.center);
	\end{pgfonlayer}
\end{tikzpicture}
	\caption{Barycentric vs.\ chromatic subdivision.}
	\label{fig:subdivs}
\end{figure}

Since we are sometimes interested in relating combinatorial topology with its point-set topology counterpart, we need a precise definition for the topological spaces represented by abstract simplicial complexes. We start with the geometric interpretation of a simplex.

\begin{definition}[Geometric simplex] \label{def:geosimp}
	Let $\mathcal{K}$ be an abstract simplicial complex, and $\sigma \in F(\mathcal{K})$.
	
	We define the geometric realization of $\sigma$, denoted by $| \sigma |$ as the set of affine functions from $\sigma$ into the closed interval $[0,1]$. That is $| \sigma | =  \{ f : \sigma \rightarrow [0,1] \; | \; \sum \limits_{v \in \sigma} f(v) = 1\}$. Elements of $| \sigma |$ are called points of $|\sigma|$.
	
	In order to make proofs more readable, we will also express the points of a geometric simplex in algebraic form, namely, as $\sum \limits_{v \in \sigma} f(v) \cdot v$.
\end{definition}

Intuitively, the affine functions form a coordinate system, where the vertices in $\sigma$ are the generators. A $k$-dimensional geometric simplex is homeomorphic to a $k$-dimensional ball, and can be embedded into a $k$-dimensional Euclidean space. Notice that the above definition has the advantage ob being independent from embeddings, however.

While we have defined the geometric realization of an individual simplex, we also need a point-set topological space for the whole abstract simplicial complex.

\begin{definition}[Geometric Simplicial Complex]\label{def:geosimpcomp}
	For an abstract simplicial complex $\mathcal{K}$, we define the geometric realization of $\mathcal{K}$, denoted by $|\mathcal{K}|$, as the quotient space $\coprod \limits _{\sigma \in F(\mathcal{K})} |\sigma| / R$. Herein, $\coprod \limits _{\sigma \in F(\mathcal{K})} |\sigma|$ is the disjoint union of all geometric simplices of $\mathcal{K}$, and $R$ is the equivalence relation given by $f \underset{R}{\sim} g$ iff $f^{-1}((0,1]) = g^{-1}((0,1])$ and $f(v) = g(v)$ for all $v \in f^{-1}((0,1])$.
\end{definition}

Via this definition, we create disjoint geometric simplices and glue them along shared points. Observe that the equivalence relation $R$ defines shared points as functions that have the same non-zero coordinates.

We note that a geometric simplicial complex $| \mathcal{K} |$ is homeomorphic to both its barycentric subdivision $|\textrm{Bary}(\mathcal K)|$ and to its chromatic subdivision $| \textrm{Ch}(\mathcal{K})|$. Therefore, it is also homeomorphic to any finitely iterated subdivision.

\begin{definition}[Geometric Star of a Vertex]\label{def:star}
	Let $v \in V(\mathcal{K})$ be a vertex of an abstract simplicial complex $\mathcal{K}$. We define the closed star of $v$, $\cstar(v)$ as the closed subset of $ \vert \mathcal{K} \vert$ induced by the geometric realization of the simplices that contain $v$. More precisely, $\cstar(v) = \bigcup \limits _{v \in \sigma } \vsigma$.
	The open star of a vertex is the interior of the closed star, and denoted by $\ostar(v) = \interior(\cstar(v))$; note that $\interior(v)=v$ for every vertex. Unless explicitly stated, we will refer to the closed star of a vertex $v$ when using $\cstar(v)$.
\end{definition}

\subsection{The Topological Model}
\label{sec:distcomp}

In the previous subsection, we provided a reasonably comprehensive list of basic definitions of combinatorial topology. In this section, we will use and appropriately extend them in order to be able to model the wait-free asynchronous shared memory (ASM) model. We will focus on the \emph{Iterated Immediate Snapshot} (IIS) Model described in \cite{HKR13}, since it is conveniently equivalent to other wait-free shared memory models in terms of computability, and it relates directly to the topological framework.

First, we need to model input and output configurations for distributed computations. This will be done by dedicated simplicial complexes. In the input complex, local input states are modeled as vertices, and global initial configurations as faces. Since we are interested in colored tasks, processes are labeled by their unique ids, which is modeled by a proper coloring of the vertices.

\begin{definition}[Input complex]
	We define the input complex of a distributed system, denoted by $\mathcal{I}$, as the following abstract simplicial complex:
\begin{align}	
	V(\I) &= \{(p_i,v_i) \; | \; p_i \textrm{ is a process id, and } v_i \textrm{ is a valid input value for }p_i \}, \nonumber\\
	F(\I) &= \{ (w_1, \ldots, w_n) \; | \; \textrm{ there is a valid global input configuration that is consistent with all }w_i \}.\nonumber
\end{align}
\end{definition}

We can use a symmetrical definition for an output complex.

\begin{definition}[Output complex]
	We define the output complex of a distributed system, denoted by $\O$, as the following abstract simplicial complex:
\begin{align}		
	V(\O) &= \{(p_i,v_i) \; | \; p_i \textrm{ is a process id, and } v_i \textrm{ is a valid output value for }p_i \}, \nonumber\\
	F(\O) &= \{ (w_1, \ldots, w_n) \; | \; \textrm{ there is a valid global output configuration that is consistent with all }w_i \}.\nonumber
\end{align}	
\end{definition}

Carrier maps are used to specify the validity conditions of a task. A simplex in the input complex $\mathcal{I}$, which corresponds to a valid initial configuration, is mapped to a subcomplex of the output complex $\mathcal{O}$. This subcomplex corresponds to the valid different outputs that are allowed for this input. 

\begin{definition}[Carrier Set]\label{def:carrset}
	For a point $x \in \vert \mathcal{K} \vert$, let the carrier of $x$, denoted $\sigma = \textrm{carr}(x, \mathcal{K})$, be the unique smallest simplex $\sigma \in \mathcal{K}$ such that $x \in \vert \sigma \vert$. We can also define the carrier of a set $S \subseteq \vert \mathcal{K} \vert, \textrm{ as } \textrm{carr}(S,\mathcal{K}) = \displaystyle \bigcup\limits_{x \in S} \textrm{carr}(x, \mathcal{K})$.
\end{definition}

Notice that any chromatic subdivision $\textrm{Sub}(\mathcal{I})$ of a chromatic simplicial complex $\mathcal{I}$ induces a carrier map $\Phi: \mathcal{I} \rightarrow \textrm{Sub}(\mathcal{I})$. $\Phi$ maps each simplex $\sigma$ into a simplicial complex $\textrm{Sub}(\sigma)$. 

The following definition gives the standard meaning of the carrier $\textrm{carr}(\tau, \Phi)$
of a simplex $\tau \in \mathcal{O}$ for a strict carrier map $\Phi: \mathcal{I} \rightarrow 2^{\mathcal{O}}$:

\begin{definition}\label{def:carrcomp}
	Let $ \mathcal{I} , \mathcal{O}$ be finite simplicial complexes and $\Phi : \mathcal{I} \rightarrow 2^{\mathcal{O}}$ be a strict carrier map. For each simplex $\tau \in \Phi(\mathcal{I})$, there exists a unique simplex $\sigma \in \mathcal{I}$ of smallest dimension, such that $\tau \in \Phi(\sigma)$. We say that this $\sigma$ is the carrier of $\tau$ under $\Phi$, denoted $\textrm{carr}(\tau, \Phi)$.
\end{definition}

A carrier map $\Delta: \I \rightarrow 2^{F(\O)}$ can be used to specify the validity conditions for a distributed task, i.e., which outputs are valid for each particular input. 

\begin{definition}[Task]
	A distributed task $T$ is a triple $\langle \I, \O, \Delta \rangle$, where $\I$ is a pure abstract simplicial complex, $\O$ is a pure abstract simplicial complex of the same dimension as $\I$, and $\Delta$ is a carrier map.
\end{definition}

We have now developed a topological model that captures the input and output conditions for a distributed system. However, we are also interested in the evolution of the system throughout an execution. This is modeled by another simplicial complex, the protocol complex. 

\begin{definition}[Protocol complex]
	We define the protocol complex of a distributed system, denoted by $\mathcal{P}$, as the following abstract simplicial complex:
\begin{align}
	V(\mathcal{P}) &= \{(p_i, s) \; | \; p_i \textrm{ is a process id, and } s \textrm{ is a valid local state for } p_i \},\nonumber\\
	F(\mathcal{P}) &= \{ (w_1, \ldots, w_n) \; | \; \textrm{ there is a valid global state configuration that is consistent with all }w_i \}.\nonumber
\end{align}
\end{definition}

In contrast to the input and output complexes, which are very well defined, there is no standard way of describing the protocol complex, as it intrinsically depends on the peculiarities of the computing model, like message passing versus shared memory communication, asynchronous versus synchronous execution etc. In fact, it is the communication between processes over time that determines the
evolution of the protocol complex, which starts out from the input complex.

In the IIS model, all protocol complexes are chromatic subdivisions of $\mathcal{I}$. More specifically, an execution with $k$ communication rounds is represented by the the k-th chromatic subdivision of $\I$, that is, $\mathcal{P}=\Ch^k(\I)$.

We can now finally state what it means to solve a task in the topological model.

\begin{definition}[Task solvability]
	We say that a task $T = \langle \I, \O, \Delta \rangle$ is solvable in a distributed model $\mathcal{M}$, if there exists a protocol complex $\mathcal{P}$ associated to the model $\mathcal{M}$ and a chromatic simplicial map $\mu: \mathcal{P} \rightarrow \O$ that is carried by $\Delta$.
\end{definition}

\end{toappendix}

\section{Continuous Task Solvability in the ASM Model}
\label{sec:results}

In this section, we show that chromatic functions precisely capture the notion of computability 
for the wait-free asynchronous read/write shared memory (ASM) model. Our results hence indeed
provide an alternative proof of the ACT \cref{th:ACTif}, by means of a simple reduction based on \cref{lemma:indtask}.

\begin{lemma} \label{lemma:indtask}
A continuous task $T = \langle \mathcal{I}, \mathcal{O},f \rangle$ has a solution in ASM if and only if its induced task $T_f = \langle \mathcal{I}, \mathcal{O},\Delta_f \rangle$ has a solution in ASM.
\end{lemma}

\begin{proof}
Assume that  a continuous task $T = \langle \mathcal{I}, \mathcal{O}, f \rangle$ has a solution, that is, that there exists a subdivision $\Sub(\mathcal{I})$ and a decision map $\mu: \Sub(\mathcal{I}) \rightarrow \mathcal{O}$ such that for each $\sigma \in \Sub(\mathcal{I}), \mu(\sigma) \subseteq  \carr(f (\vsigma),\mathcal{O}) $. It follows from \cref{def:indtask} that $\mu$ solves $T_f$.

Conversely, assume that the induced task $T_f$ has a solution. According to the ACT \cref{th:ACTif}, there exists a subdivision $\Sub(\mathcal{I})$ and a decision map $\mu : \Sub(\mathcal{I}) \rightarrow \mathcal{O}$ carried by $\Delta_f$. From \cref{def:tassol} of the carrier $\Delta_f$ of $T_f$, it follows that for any $\sigma \in \Sub(\mathcal{I})$, $\mu(\sigma) \subseteq \carr(f(\vsigma), \mathcal{O})$. Therefore, by \cref{def:tassol}, $\mu$ solves $T$.
\end{proof}

The following \cref{thm:asmeq} establishes the equivalence of ASM task solvability and
the existence of a chromatic task.

\begin{theorem}[CACT]
 \label{thm:asmeq}
A task $\langle \mathcal{I}, \mathcal{O}, \Delta \rangle$ is solvable in ASM if and only if there exists a continuous task $\langle \mathcal{I}, \mathcal{O}, f \rangle$ such that $f(\vsigma) \subseteq \vert \Delta (\sigma) \vert$ for any input simplex $\sigma \in \mathcal{I}$.
\end{theorem}

\begin{proof}
If $\langle \mathcal{I}, \mathcal{O}, \Delta \rangle$ is solvable, then there exists a subdivision $\Sub(\mathcal{I})$ and a decision map $\mu : \Sub(\mathcal{I}) \rightarrow \mathcal{O}$ carried by $\Delta$. Since $\mu$ is a chromatic simplicial map, \cref{lemma:chrmapcont} reveals that its geometric realization $\vmu: \vI \rightarrow \vO$ is a chromatic function, which satisfies
$\vmu(\vsigma) \subseteq \vert \Delta (\sigma) \vert$ for any input simplex $\sigma \in \mathcal{I}$.

Conversely, let $T = \langle \mathcal{I}, \mathcal{O}, f \rangle$ be a continuous task such that $f(\vert \sigma \vert ) \subseteq \vert \Delta (\sigma) \vert$, and consider the task $T_f = \langle \mathcal{I}, \mathcal{O}, \Delta \rangle$. Since $f$ is chromatic, it follows from \cref{thm:chrappr} that $f$ has a chromatic approximation $\mu_f  : \Sub(\mathcal{I}) \rightarrow \mathcal{O}$. Since $f(\vsigma) \subseteq \vert \Delta (\sigma) \vert$, \cref{def:tassol} implies that $\mu_f$ solves the induced task $T_f$.
\end{proof}

Finally, the following \cref{cor:main} shows that every continuous task can be solved in the ASM. Together with \cref{thm:asmeq}, it provides the continous counterpart of the only-if 
direction of the ACT \cref{th:ACTif}.

\begin{theorem}
\label{cor:main}
Any continuous task $T=\langle \mathcal{I}, \mathcal{O},f \rangle$ is solvable in ASM.
\end{theorem}

\begin{proof}
Let $T = \langle \mathcal{I}, \mathcal{O},f \rangle$ be a continuous task. It follows from \cref{thm:chrappr} that $f$ has a chromatic approximation. According to \cref{def:tassol}, it therefore has a solution in ASM. \cref{lemma:indtask} thus confirms that the induced task $T_f = \langle \I, \O, \Delta_f \rangle$ also has solution in ASM.
\end{proof}

\section{Application Example: Consensus-Preferent Approximate Agreement}
\label{sec:app}

As we have shown, chromatic functions precisely characterize task solvability under the wait-free asynchronous shared memory model. However, their expressive power goes way beyond that, opening up interesting future research areas. We will demonstrate this by using a continuous task to specify preferences for particular output configurations. More specifically, 
we consider a two-process system in ASM and specify a $1/3$ binary approximate agreement task with a fixed \emph{preference} $0<K<1$ for \emph{exact} agreement. The parameter $K$ determines the fraction of all the executions of the system that
will terminate in a configuration where all processes decide on the same value. For example, if one assumes that all 
IIS runs are equally likely, then $K$ gives the probability that a randomly chosen run terminated with exact agreement.

Since every execution corresponds to a
simplex in the chromatic subdivision $\Sub(\I)$ guaranteed by \cref{thm:chrappr}, this preference constraint can be easily expressed by means of piecewise linear functions in a continuous task specification. More specifically, we will define a chromatic function $f: \vert \mathcal{I} \vert \rightarrow \vert \mathcal{O} \vert$ for the task at hand as follows (see \cref{fig:prefagr} for an illustration).

\begin{figure}[h!]
\centering
\scalebox{0.7}{	\begin{tikzpicture}
	\begin{pgfonlayer}{nodelayer}
		\node [style=VG] (0) at (-9, 0) {};
		\node [style=VR] (1) at (0, 0) {};
		\node [style=VG] (2) at (0, 9) {};
		\node [style=VR] (3) at (-9, 9) {};
		\node [style=none] (4) at (-9, -1) {$p_1, 0$};
		\node [style=none] (5) at (0, -1) {$p_2, 0$};
		\node [style=none] (6) at (0, 10) {$p_1, 1$};
		\node [style=none] (7) at (-9, 10) {$p_2, 1$};
		\node [style=VG] (8) at (9, 0) {};
		\node [style=VR] (9) at (18, 0) {};
		\node [style=VG] (10) at (18, 9) {};
		\node [style=VR] (11) at (9, 9) {};
		\node [style=none] (12) at (9, -1) {$p_1, 0$};
		\node [style=none] (13) at (18, -1) {$p_2, 0$};
		\node [style=none] (14) at (18, 10) {$p_1, 1$};
		\node [style=none] (15) at (9, 10) {$p_2, 1$};
		\node [style=VG] (16) at (9, 6) {};
		\node [style=VG] (17) at (18, 3) {};
		\node [style=VR] (18) at (9, 3) {};
		\node [style=VR] (19) at (18, 6) {};
		\node [style=none] (20) at (7, 3) {$p_2, 1/3$};
		\node [style=none] (21) at (20, 6) {$p_2, 2/3$};
		\node [style=none] (22) at (20, 3) {$p_1, 1/3$};
		\node [style=none] (23) at (7, 6) {$p_1, 2/3$};
		\node [style=none] (24) at (-5, -2) {$\mathcal{I}$};
		\node [style=none] (25) at (14, -2) {$\mathcal{O}$};
		\node [style=VBfaded] (26) at (-9, 4) {};
		\node [style=VBfaded] (27) at (-9, 1) {};
		\node [style=none] (28) at (-10, 4) {$x_2$};
		\node [style=none] (29) at (-10, 1) {$x_1$};
		\node [style=none] (30) at (-4.5, 8.25) {$\sigma_1$};
		\node [style=none] (31) at (-4.75, -0.75) {$\sigma_0$};
		\node [style=none] (32) at (-7.25, 4.5) {$\sigma_2$};
		\node [style=none] (33) at (-0.75, 4.5) {$\sigma_3$};
		\node [style=none] (34) at (13.5, -0.5) {$\tau_0$};
		\node [style=none] (35) at (13.5, 9.5) {$\tau_1$};
		\node [style=VBfaded] (36) at (-9, 5) {};
		\node [style=none] (37) at (-10, 5) {$x_3$};
		\node [style=VBfaded] (38) at (-9, 8) {};
		\node [style=none] (39) at (-10, 8) {$x_4$};
	\end{pgfonlayer}
	\begin{pgfonlayer}{edgelayer}
		\draw (0) to (1);
		\draw (1) to (2);
		\draw (2) to (3);
		\draw (8) to (9);
		\draw (10) to (11);
		\draw [style=EdgeBlue, in=270, out=90] (0) to (27);
		\draw [style=EdgeBlue] (8) to (18);
		\draw [style=EdgeRed] (27) to (26);
		\draw [style=EdgeRed] (18) to (17);
		\draw [style=EdgeBlue] (17) to (19);
		\draw [style=EdgeBlue] (26) to (36);
		\draw [style=EdgeRed] (36) to (38);
		\draw [style=EdgeBlue] (38) to (3);
		\draw [style=EdgeRed] (19) to (16);
		\draw [style=EdgeBlue] (16) to (11);
		\draw (18) to (16);
		\draw (9) to (17);
		\draw (19) to (10);
	\end{pgfonlayer}
\end{tikzpicture}}
	\caption{The chromatic function for the consensus-preferent 1/3 approximate agreement task.
It maps a portion of $K$ of the left edge in the input simplex $\I$, represented by the red segments $(x_1,x_2)$ and $(x_3,x_4)$, to the red horizontal edges in the middle of $\O$,
and the remaining blue segments to the blue vertical edges on the left and right of $\O$.}
	\label{fig:prefagr}
\end{figure}

Let $\sigma_0$ be the simplex of $\mathcal{I}$ that corresponds to the input configuration $\{ (p_1,0) (p_2,0) \}$, and $\tau_0$ the simplex of $\mathcal{O}$ that corresponds to the output configuration $\{(p_1,0), (p_2,0)\}$. For $x \in \vert \sigma_0 \vert = \lambda \cdot (p_1,0) + (1-\lambda) \cdot (p_2,0)$ for $0\leq \lambda \leq 1$, we define $f(x) = \lambda \cdot (p_1,0) + (1-\lambda)\cdot (p_2,0) \in \vert \mathcal{O} \vert $. We extend this definition to $\sigma_1 = \{ (p_1,1), (p_2,1)\} \in \mathcal{I}$ and $\tau_1 =\{ (p_1,1), (p_2,1)\} \in \mathcal{O}$ in a completely analogous way. This corresponds to mapping the top resp.\ bottom edge of the input complex to the top resp.\ bottom edge of the output complex.

In order to complete our definition of $f$, we also need to define it on $\sigma_2 = \{ (p_1,0), (p_2,1)\}$ and on $\sigma_3 = \{(p_1,1), (p_2,0)\}$. We describe our construction for $f$ restricted to $\vert \sigma_2 \vert$; the construction for $f$ on $\vert \sigma_3 \vert$ is completely analogous. Let $M_1$ be any positive number such that $0<K<M_1<1$, and set $R = (1-M_1)/3$. Consider $x_1,\dots,x_4 \in \vert \sigma_2 \vert$ defined by $x_1 = R \cdot (p_1,0) + (M_1 + 2R) \cdot(p_2,1)$, $x_2 = (M_1/2+R) \cdot (p_1,0) +  (2R+ M_1/2) \cdot(p_2,1)$, $x_3= (2R + M_1/2) \cdot (p_1,0) + (M_1/2 + R) (p_2,1)$, and $ x_4 = (2R + M_1) \cdot (p_1,0)+ R \cdot(p_2,1)$.
We now define $f$ on $\vert \sigma_2 \vert $ as a piecewise affine function that maps segment $[(p_1,0), x_1]$ to segment $[(p_1,0),(p_2,1/3)]$. More specifically, for $x$ in segment $[(p_1,0), x_1]$ given by $x = \lambda \cdot (p_1,0) + (1-\lambda) \cdot x_1$ for some $0 \leq \lambda\leq 1$, we define $f(x) = \lambda \cdot (p_1,0) + (1-\lambda) \cdot (p_2,1/3)$. In the same way, we can map segments $[x_1,x_2]$ to $[(p_2,1/3), (p_1,1/3)]$, $[x_2,x_3]$ to $[(p_1,1/3), (p_2,2/3)]$, $[x_3,x_4]$ to $[(p_2,2/3),(p_1,2/3)]$ and $[x_4,(p_2,1)]$ to $[(p_1,2/3),(p_2,1)]$.
 
Since it is obvious that the so-constructed $f$ is a chromatic function, \cref{thm:asmeq} 
guarantees that this task is solvable in the wait-free ASM model. More specifically, we obtain
the following \cref{thm:cpapproxagreement}:

 \begin{theoremrep}\label{thm:cpapproxagreement}
Let $\textrm{Sub}(\mathcal{I})$ be a chromatic subdivision of $\mathcal{I}$, such that each facet has the same size less than $(M_1 - K)/4$, and let $\mu$ be the chromatic approximation to $f$ with respect to  $\Sub(\mathcal{I})$ guaranteed by \cref{thm:chrappr}. Then, $\mu$ can be used to solve $1/3$-approximate agreement with a proportion of at least $K$ executions that output the same value in the wait-free ASM. In the IIS model, $r=\log_3(4/(M_1-K))$ communication rounds 
are enough for solving this task.
 \end{theoremrep}

\begin{proof}
From the definition of $f$, it is apparent that the segments ${[x_1,x_2]}$ and ${[x_3,x_4]}$ both have length $M/2$. Since they are mapped to the red segments in in $\O$ in \cref{fig:prefagr}, it follows that if $\sigma \in \textrm{Sub}(\I)$, and $| \sigma | \subseteq \textrm{int}({[x_1,x_2]})$, then $\mu (\sigma) = (p_1,1/3), (p_2,1/3)$, and if $| \sigma | \subseteq \textrm{int}({[x_3,x_4]})$, then  $\mu (\sigma) = (p_1,2/3), (p_2,2/3)$. Since ${[x_1,x_2]}$ has only two border points, it must have at least $\lceil{(M_1 / 2) / D} \rceil - 2$ internal facets, where $D$ is the facet size. Analogously, ${[x_3,x_3]}$ must also have at least  $\lceil{(M_1 / 2) / D} \rceil - 2$ internal facets. In total, we have at least $M_1 /D - 4$ facets in the original $\sigma_2$ that map to an exact consensus output configuration, out of a total of $1/D$ facets. Thus, the proportion of the subdivided simplices that results in exact agreement is at least $M_1 - 4 \cdot D$. From the statement of our theorem, we know that $D \leq (M_1 - K)/4$, therefore the proportion is at least $M_1 -(M_1-K) = K$.

Exactly the same reasoning applies to $\sigma_3$, which also guarantees exact agreement for a proportion at least $K$. On the other hand, all output configurations reachable from $\sigma_0$ and $\sigma_1$ guarantee consensus, therefore their proportion is 1. Since $\sigma_0, \sigma_1, \sigma_2$, and $\sigma_3$ are all the facets of $\I$, and since the subdivision is uniform, it follows that the proportion of the facets of $\textrm{Sub}(\I)$ leading to exact consensus is at least $K$.

Since the standard chromatic subdivision divides a 2-dimensional simplex uniformly and reduces its size to $1/3$ of the original size, \cref{thm:cpapproxagreement} implies that $r=\log_3(4/(M_1-K))$ communication rounds 
are indeed enough for solving this task in the IIS model.
\end{proof}

\begin{toappendix}
\begin{figure}[h!]
\centering
\scalebox{0.7}{\begin{tikzpicture}
	\begin{pgfonlayer}{nodelayer}
		\node [style=VG] (0) at (-9, 3) {};
		\node [style=VR] (1) at (0, 3) {};
		\node [style=VG] (2) at (0, 12) {};
		\node [style=VR] (3) at (-9, 12) {};
		\node [style=none] (4) at (-9, 2) {$p_1, 0$};
		\node [style=none] (5) at (0, 2) {$p_2, 0$};
		\node [style=none] (6) at (0, 13) {$p_1, 1$};
		\node [style=none] (7) at (-9, 13) {$p_2, 1$};
		\node [style=VG] (8) at (9, 0) {};
		\node [style=VR] (9) at (18, 0) {};
		\node [style=VG] (10) at (18, 15) {};
		\node [style=VR] (11) at (9, 15) {};
		\node [style=none] (12) at (9, -1) {$p_1, 0$};
		\node [style=none] (13) at (18, -1) {$p_2, 0$};
		\node [style=none] (14) at (18, 16) {$p_1, 1$};
		\node [style=none] (15) at (9, 16) {$p_2, 1$};
		\node [style=VG] (16) at (9, 6) {};
		\node [style=VG] (17) at (18, 3) {};
		\node [style=VR] (18) at (9, 3) {};
		\node [style=VR] (19) at (18, 6) {};
		\node [style=none] (20) at (7, 3) {$p_2, 1/3$};
		\node [style=none] (21) at (20, 6) {$p_2, 2/3$};
		\node [style=none] (22) at (20, 3) {$p_1, 1/3$};
		\node [style=none] (23) at (7, 6) {$p_1, 2/3$};
		\node [style=none] (24) at (-5, 1) {$\mathcal{I}$};
		\node [style=none] (25) at (14, -2) {$\mathcal{O}$};
		\node [style=VBfaded] (26) at (-9, 7) {};
		\node [style=VBfaded] (27) at (-9, 4) {};
		\node [style=none] (30) at (-4.5, 11.25) {$\sigma_1$};
		\node [style=none] (31) at (-4.75, 2.25) {$\sigma_0$};
		\node [style=none] (32) at (-7.25, 7.5) {$\sigma_2$};
		\node [style=none] (33) at (-0.75, 7.5) {$\sigma_3$};
		\node [style=none] (34) at (13.5, -0.5) {$\tau_0$};
		\node [style=none] (35) at (13.5, 15.75) {$\tau_1$};
		\node [style=VBfaded] (36) at (-9, 8) {};
		\node [style=VBfaded] (38) at (-9, 11) {};
		\node [style=VR] (40) at (9, 9) {};
		\node [style=VG] (41) at (9, 12) {};
		\node [style=VR] (42) at (18, 12) {};
		\node [style=VG] (43) at (18, 9) {};
		\node [style=none] (44) at (7, 9) {$p_2, 1/3$};
		\node [style=none] (45) at (7, 12) {$p_1, 2/3$};
		\node [style=none] (46) at (20, 9) {$p_1, 1/3$};
		\node [style=none] (47) at (20, 12) {$p_2, 2/3$};
		\node [style=VBfaded] (48) at (-9, 9.25) {};
		\node [style=VBfaded] (49) at (-9, 10) {};
		\node [style=VBfaded] (50) at (-9, 5) {};
		\node [style=VBfaded] (51) at (-9, 5.75) {};
	\end{pgfonlayer}
	\begin{pgfonlayer}{edgelayer}
		\draw (0) to (1);
		\draw (1) to (2);
		\draw (2) to (3);
		\draw (8) to (9);
		\draw (10) to (11);
		\draw [style=EdgeBlue, in=270, out=90] (0) to (27);
		\draw [style=EdgeBlue] (8) to (18);
		\draw [style=EdgeRed] (18) to (17);
		\draw [style=EdgeBlue] (17) to (19);
		\draw [style=EdgeBlue] (26) to (36);
		\draw [style=EdgeBlue] (38) to (3);
		\draw [style=EdgeRed] (19) to (16);
		\draw (18) to (16);
		\draw (9) to (17);
		\draw [style=EdgeBlue] (16) to (40);
		\draw [style=EdgeBlue] (43) to (42);
		\draw [style=EdgeBlue] (41) to (11);
		\draw [style=Enone] (40) to (41);
		\draw [style=Enone] (19) to (43);
		\draw [style=Enone] (42) to (10);
		\draw [style=EdgeRed] (40) to (43);
		\draw [style=EdgeRed] (41) to (42);
		\draw [style=EdgeBlue] (48) to (49);
		\draw [style=EdgeBlue] (50) to (51);
		\draw [style=EdgeRed] (38) to (49);
		\draw [style=EdgeRed] (48) to (36);
		\draw [style=EdgeRed] (26) to (51);
		\draw [style=EdgeRed] (50) to (27);
	\end{pgfonlayer}
\end{tikzpicture}}
	\caption{The chromatic function for and extended version of the consensus-preferent 1/3 approximate agreement task. In contrast to \cref{fig:prefagr}, it allows both processes to decide on $1/3$ resp.\ $2/3$ when the other process decides on $2/3$ resp.\ $1/3$.}
	\label{fig:prefagrsymm}
\end{figure}
\end{toappendix}

We conclude this section by noting that the above example also reveals 
that continuous tasks
allow for more natural fine-grained specifications than standard tasks:
It is apparent from \cref{fig:prefagr}  that the continuous task specification
does not allow a configuration where $(p_1,p_2)$ have decided $(2/3,1/3)$ when
starting from the input $(0,1)$ in $\sigma_2$, which would be allowed by the 
standard (colorless) task specification of $1/3$ approximate agreement. To also allow
this behavior, one could use the continuous task specification illustrated in
\cref{fig:prefagrsymm} in the appendix, however.

\section{Conclusions and Future Work}
\label{sec:concl}

In this paper, we defined chromatic functions as continuous functions with some color preservation properties, and showed that they precisely characterize task solvability in a wait-free asynchronous shared memory model (ASM). Chromatic functions can be seen as the continuous analogue of chromatic simplicial maps, and  provide a means for expressing refined task specifications, e.g., for resolving non-determinism. Overall, they provide a purely topological formalization of the computability power of ASM models. 

Technically, our results rest on the novel notion of  chromatic approximations, a chromatic analog of the well-known simplicial approximation theorem.
The main feature guaranteed by chromatic functions is preserving the color structure, which is not native to point-set topology. In fact, chromatic functions are a formalization of quite intuitive rules of what can and can't be done to transform an input complex into an output complex:  For example, stretching and even possibly folding over along an edge or a facet is permitted. Puncturing or cutting would violate continuity and is hence forbidden, however. Less obvious is the fact that any collapsing is prohibited, since this would imply a violation to the color structure. 

The utility of continuous tasks is not limited to standard task solvability. We demonstrated this fact by adding density constraints for the outputs to the classic 1/3 approximate agreement task, namely, a lower bound on the fraction of executions that actually guarantee exact consensus. 
This gives a guarantee that exact agreement will happen frequently enough on average.

Regarding future work, our results open up promising research avenues in several directions. 
Our perspective might be useful to derive characterizations of task solvability in
other models of computation, especially those for which there exist no topological characterization; one example are models where communication is performed via shared objects used in practice, which are more powerful than read/write operations.
Besides the power of continuous tasks for refining task specifications, we also believe that the ability to incorporate density constraints is interesting from a non-worst-case quality-of-service perspective in general. Last but not least, we expect that continuous tasks with density constraints could also be useful for characterizing non-terminating tasks, such as stabilizing ones, and asymptotic ones~\cite{FNS18:PODC}. This is because task solvability is expressed only in terms of geometric simplicial complexes, which may allow to incorporate output sequences via continuous functions defined on the output complex.


\section{Discussion of additional related work}
\label{sec:relatedWork}

\textbf{Alternative characterization approaches:}
Most alternative/generalized characterizations we are aware of either refer to \cite{GKM14:PODC} or adapt
\cite[Ch.~11]{HKR13} the original construction \cite{HS99:ACT} of a chromatic carrier-preserving simplicial decision map
from a ``perturbation-based'' construction of a chromatic subdivision of $\mathcal{I}$, which
requires the output complex resp.\ the carrier map to be link-connected. In particular, in \cite{GKM14:PODC}, 
Gafni, Kuznetsov and Manolescu provide a generalized ACT for general 
Sub-IIS models, leading to affine tasks~\cite{Kuznetsov12}, which also include non-compact models like $t$-resilient and obstruction-free ones.
Their GACT depends on the existence of a chromatic simplicial map, however. 
In the wait-free case \cite[Cor.~7.1]{GKM14:PODC}, the authors just refer to
the construction in \cite{HS99:ACT}. For general models, a dedicated 
\cite[Thm.~8.4]{GKM14:PODC} is provided. However, it requires a link-connected
output complex and just resorts to \cite[Lem.~4.21, Thm.~5.29]{HS99:ACT} in its proof
as well.

In~\cite{SaraphHG18}, Saraph, Herlihy and Gafni explain that
\emph{``... the ACT contains one difficult step. Using the classical simplicial approximation theorem, it is straightforward to construct a simplicial map having all desired properties except that
of being color-preserving. To make this map color-preserving required a rather long construction employing mechanisms from point-set topology, such as balls and Cauchy sequences.''}
In their paper, they give an alternative proof strategy for the ACT, in which the essential
chromatic property is guaranteed by a distributed convergence algorithm, 
rather than by a combinatorial construction, whose proof is quite subtle and long.
They observe that the convergence algorithm
can be applied to more general continuous functions from $\mathcal{I}$ to $\mathcal{O}$,
carried by $\Delta$, given the assumption that $\Delta(\sigma)$ is link connected
for all $\sigma$ in $\mathcal{I}$. This implies \cite[Theorem 6.1]{SaraphHG18}, which
states that if there is a continuous map $f:\vI \to \vO$ that is carried by
a link-connected carrier map $\Delta$, then there exists a carrier-preserving simplicial map
from a chromatic subdivision of $\mathcal{I}$ to $\mathcal{O}$. We note, however, that
this is only a sufficient condition, but nothing is stated about the other direction:
Whereas constructing a continous map from a carrier-preserving
simplicial map is easy, it is not clear how to enforce link-connectivity of $\Delta$.

In fact, link-connectivity is a property required from the output complex resp.\
the images of the carrier map $\Delta$, and it is easy to find examples (like
the Hourglass task discussed in \cref{sec:intro}) of continuos tasks that are not 
link-connected. In particular, the assumption that $\Delta(\sigma)$ is link connected 
can be violated even by admitting just one additional possible output configuration $\rho \in \Delta(\sigma)$ (e.g.,
involving a default value $\bot$ as in quasi-consensus \cite{GR06:TCJ})
for just one simplex $\sigma\in\I$. 

 \begin{wrapfigure}[7]{r}{0.25\textwidth}
 \centering
 \vspace{-0.9cm}
\scalebox{1.0}{\begin{tikzpicture}
	\begin{pgfonlayer}{nodelayer}
		\node [style=none] (38) at (-7.25, 11.75) {};
		\node [style=none] (39) at (-7.25, 11.75) {};
		\node [style=none] (40) at (-7.25, 11.75) {};
		\node [style=VGsmall] (41) at (-4, 11.75) {};
		\node [style=VRsmall] (42) at (-6, 10.75) {};
		\node [style=none] (43) at (-6, 10.75) {};
		\node [style=none] (44) at (-4, 11.75) {};
		\node [style=none] (47) at (-6, 11.25) {1};
		\node [style=none] (48) at (-4, 12.25) {1};
		\node [style=none] (49) at (-7, 12.25) {1};
		\node [style=VYsmall] (50) at (-7.25, 11.75) {};
		\node [style=VGsmall] (51) at (-4, 7.75) {};
		\node [style=VRsmall] (52) at (-6.5, 7) {};
		\node [style=VYfaded] (53) at (-7.5, 8) {};
		\node [style=none] (54) at (-7.5, 8) {};
		\node [style=none] (55) at (-6.5, 7) {};
		\node [style=none] (56) at (-4, 7.75) {};
		\node [style=VYsmall] (57) at (-7.5, 8) {};
		\node [style=none] (58) at (-6.5, 7) {};
		\node [style=none] (59) at (-6.5, 7) {};
		\node [style=none] (60) at (-6.5, 7) {};
		\node [style=none] (61) at (-4, 7.75) {};
		\node [style=none] (62) at (-7.5, 7.5) {0};
		\node [style=none] (63) at (-6.5, 6.5) {0};
		\node [style=none] (64) at (-4, 7.25) {0};
		\node [style=none] (65) at (-6.5, 5.5) {$\mathcal{O}$};
		\node [style=VGsmall] (66) at (-6.75, 9.5) {};
		\node [style=none] (67) at (-7.5, 8) {};
		\node [style=none] (68) at (-6.5, 7) {};
		\node [style=none] (69) at (-6.75, 9.5) {};
		\node [style=none] (70) at (-7.25, 11.75) {};
		\node [style=none] (71) at (-6.75, 9.5) {};
		\node [style=none] (72) at (-6, 10.75) {};
		\node [style=VYsmall] (73) at (-5, 9.5) {};
		\node [style=none] (74) at (-5, 9.5) {};
		\node [style=none] (75) at (-6.5, 7) {};
		\node [style=none] (76) at (-4, 7.75) {};
		\node [style=none] (77) at (-6, 10.75) {};
		\node [style=none] (78) at (-5, 9.5) {};
		\node [style=none] (79) at (-4, 11.75) {};
		\node [style=VRsmall] (80) at (-5.75, 10) {};
		\node [style=none] (81) at (-7.5, 8) {};
		\node [style=none] (82) at (-4, 7.75) {};
		\node [style=none] (83) at (-5.75, 10) {};
		\node [style=none] (84) at (-7.25, 11.75) {};
		\node [style=none] (85) at (-5.75, 10) {};
		\node [style=none] (86) at (-4, 11.75) {};
		\node [style=none] (87) at (-7.75, 9.75) {$\bot$};
		\node [style=none] (88) at (-4, 9.5) {$\bot$};
		\node [style=none] (89) at (-6.25, 10) {$\bot$};
	\end{pgfonlayer}
	\begin{pgfonlayer}{edgelayer}
		\draw (41) to (42);
		\draw (42) to (50);
		\draw (50) to (41);
		\draw [style=FaceYellow] (43.center)
			 to (44.center)
			 to (40.center)
			 to cycle;
		\draw [style=FaceBlue, in=135, out=45, loop] (40.center) to ();
		\draw [style=FaceBlue, in=135, out=45, loop] (40.center) to ();
		\draw [style=FaceBlue, in=135, out=45, loop] (40.center) to ();
		\draw (51) to (52);
		\draw (52) to (53);
		\draw [style=Enoned] (53) to (51);
		\draw [style=FacePur] (54.center)
			 to (55.center)
			 to (56.center)
			 to cycle;
		\draw [style=FaceBlue] (57) to (58.center);
		\draw [style=FaceBlue] (60.center) to (61.center);
		\draw (57) to (66);
		\draw (60.center) to (66);
		\draw (66) to (43.center);
		\draw (50) to (66);
		\draw [style=FaceBlue] (68.center)
			 to (69.center)
			 to (67.center)
			 to cycle;
		\draw [style=FaceBlue, in=135, out=45, loop] (67.center) to ();
		\draw [style=FaceBlue, in=135, out=45, loop] (67.center) to ();
		\draw [style=FaceBlue, in=135, out=45, loop] (67.center) to ();
		\draw [style=FaceBlue] (71.center)
			 to (72.center)
			 to (70.center)
			 to cycle;
		\draw [style=FaceBlue, in=135, out=45, loop] (70.center) to ();
		\draw [style=FaceBlue, in=135, out=45, loop] (70.center) to ();
		\draw [style=FaceBlue, in=135, out=45, loop] (70.center) to ();
		\draw (73) to (68.center);
		\draw (73) to (61.center);
		\draw [style=FaceBlue] (75.center)
			 to (76.center)
			 to (74.center)
			 to cycle;
		\draw [style=FaceBlue, in=135, out=45, loop] (74.center) to ();
		\draw [style=FaceBlue, in=135, out=45, loop] (74.center) to ();
		\draw [style=FaceBlue, in=135, out=45, loop] (74.center) to ();
		\draw (72.center) to (74.center);
		\draw (44.center) to (74.center);
		\draw [style=FaceBlue] (78.center)
			 to (79.center)
			 to (77.center)
			 to cycle;
		\draw [style=FaceBlue, in=135, out=45, loop] (77.center) to ();
		\draw [style=FaceBlue, in=135, out=45, loop] (77.center) to ();
		\draw [style=FaceBlue, in=135, out=45, loop] (77.center) to ();
		\draw [style=Enonedotted] (67.center) to (80);
		\draw [style=Enonedotted] (80) to (76.center);
		\draw [style=Enonedotted] (70.center) to (80);
		\draw [style=Enoned] (80) to (79.center);
		\draw [style=FaceBlue] (82.center)
			 to (83.center)
			 to (81.center)
			 to cycle;
		\draw [style=FaceBlue, in=135, out=45, loop] (81.center) to ();
		\draw [style=FaceBlue, in=135, out=45, loop] (81.center) to ();
		\draw [style=FaceBlue, in=135, out=45, loop] (81.center) to ();
		\draw [style=FaceBlue] (85.center)
			 to (86.center)
			 to (84.center)
			 to cycle;
		\draw [style=FaceBlue, in=135, out=45, loop] (84.center) to ();
		\draw [style=FaceBlue, in=135, out=45, loop] (84.center) to ();
		\draw [style=FaceBlue, in=135, out=45, loop] (84.center) to ();
	\end{pgfonlayer}
\end{tikzpicture}}
\end{wrapfigure} 
We illustrate this in more detail by means of a general \emph{$k$-failsafe} extension, which could be added to any conventional task specification: 
Rather than deciding
on some output value, any process, up to a maximum of $k$ processes, is also allowed to abstain. Abstaining can be formalized by outputting a distinguished value denoted by $\bot$. 
If $\mathcal{T}$ is a non-trivial task and $k<n-1$, then the output complex of the $k$-failsafe version of $T$ is not link-connected. This follows from the fact that any face $F$ containing $k$ yielding processes must be connected to any face consisting of the complementary $n-k>1$ non-yielding processes. Removing $F$ would lead to disconnected, i.e., $(-1)$-connected, facets of the original task, whereas link-connectivity would require a connectivity of $n-1-(k-1)-2 > -1$. For example, the output complex of $1$-failsafe consensus with 3 processes shown in the figure
at the left is not link connected; note that it can be viewed as three intertwined instances of the Hourglass task. In sharp contrast to link-connectivity-based characterizations, our continuous task-based
one can be applied here, and reveals that the task cannot be solved
in wait-free ASM.


We finally note that whereas it could be argued that one could 
try to restrict a non-link-connected
output complex resp.\ carrier map to a link-connected one, this is not always
the case: there are examples like the valency task introduced in \cite{ACR21:OPODIS}, where any such restriction would render an unsolvable task solvable.

%
%

\medskip

A related alternative approach uses distributed computing arguments to limit the type of
chromatic subdivisions that take place, to avoid the need for the difficult perturbation
arguments of the original ACT proof. Such an approach is described in~\cite{BGsimple97},
which uses a simulation~\cite{GafniR10simul} to the iterated model  using immediate snapshot tasks. 

Once a characterization has been proved for a specific distributed computing model,
one may use algorithmic simulations and reductions to other models.
For instance,  once the ACT has been used to characterize wait-free task solvability,
the BG simulation~\cite{BGLR01} can be used to characterize solvability when at
most $t$ processes may crash, by a distributed simulation of an algorithm
that solves the task in the other model. Whereas this simulation works only for colorless tasks (see
below),  there has also been a proposal for an extension of the BG simulation
that works for general tasks~\cite{extendedBG09}. Another example is \cite{petr2020}, which presents a
characterization of task computability in the wait-free shared-memory model in which processes, 
in addition to read/write registers, have access to $k$-test-and-set objects.
Instead of algorithmic simulations, it is also possible to directly construct reductions (maps) from a
protocol complex in one model to a protocol complex in  another model~\cite{HerlihyR12simul}.

\textbf{Colorless  computability:}
There is a  continuous characterization
of task solvability for \emph{colorless tasks}~\cite{BGLR01}, in which
we care only about the sets of input and output values, but not which processes
are associated with which values.
Many of the main tasks of interest in distributed computing are colorless: consensus, set agreement,
approximate agreement, loop agreement, etc. 
These tasks are defined by the colorless input complex $\mathcal{I}$, the
colorless output complex $\mathcal{O}$, and an input/output relation $\Delta$.
Furthermore, colorless tasks can be solved by simpler \emph{colorless algorithms},
where the processes consider the values read from the shared memory as a set,
disregarding which value belongs to which process.

The colorless ACT~\cite[Theorem 8]{HRRS17:TCS} or~\cite[Section 4.3]{HKR13} roughly says that a colorless task is wait-free solvable
if and only if  there is a continuous map $f$ from $|\mathcal{I}|$ to  
  $|\mathcal{O}|$ carried by  $\Delta$, where $|\cdot|$ denotes the geometric realization of the respective complex. Or equivalently, if and only if
  there is some $r$ and a simplicial map $\delta$
  from $Bary^r( \mathcal{I})$ to $\mathcal{O}$ carried by $\Delta$.
The proof of the colorless ACT does not require the technicalities of the original ACT result; it uses only the classic simplicial approximation theorem.

Overall, colorless computability is fairly well understood, 
because the topological techniques are simpler; the whole 
first part of the book~\cite{HKR13} is devoted to colorless tasks.
Remarkably, some of the most important wait-free results like
the set agreement impossibility and the general task undecidability~\cite{Gafni1999ThreeProcessorTA,decidHR97}
do not require the chromatic version.
Moreover, there are colorless task solvability characterizations for various other read/write models of computation, 
such as $t$-resilience~\cite{SaraphHG16}, closed adversaries~\cite{HerlihyR13topAdv} and
fair adversaries~\cite{KRH18:PODC}, even in anonymous systems where processes
have no ids~\cite{Delporte-Gallet18} and for robot coordination algorithms~\cite{AlcantaraCFR19}.

It would be interesting to  extend existing colorless results to general tasks via
our CACT, in particular, for dependent failures~\cite[Theorem 4.3]{HerlihyR10TopAdvPODC} and~\cite{HerlihyR13topAdv}, as well as for systems admitting solo executions~\cite{HRRS17:TCS} or partitions~\cite{GWSR19:SSS}. 
For solo executions, a solvability characterization for arbitrary tasks exists~\cite{soloACT2020},
but it is not continuous.

\textbf{ASM Models:}
One of the central challenges in the theory of distributed computing is determining 
the computability power of its numerous models, parameterized by types of failures 
(crash, omission, Byzantine), synchrony  (asynchronous, partially
synchronous, synchronous), and communication primitives  (message-passing,
read-write registers, powerful shared objects). For various asynchronous, synchronous
and semisynchronous models, the topology
of the protocol complex is known~\cite[Chapter 13]{HKR13}, but not a full 
  characterization in the ACT style.

The most basic models (which preserve the topology of the input complex)
are the ASM models, but even here, there
are many variants that are all equivalent with respect to task solvability. 
The original ACT result was developed in the specific model of read/write
shared memory, and was redone later in the simpler to analyze IIS model. This resulted
in another, discrete version of the ACT, justified by the existence of simulations between both models~\cite{GafniR10simul}. The book~\cite[Section 14.2]{HKR13}
describes five natural models of asynchronous computation and shows
they are all equivalent with respect to task solvability. 
The variants are based on weather processes
can take snapshots of the shared memory or just read individual registers one at a time,
and whether the shared memory can be read and written multiple times,
or just once (iterated models).
Whereas not all ASM models have exactly the same protocol complex, they have a protocol complex that
is collapsible e.g.~\cite{BenavidesR18}.

As mentioned in the introduction, instead of working in a specific read/write
wait-free model, we simply assume a model where \cref{th:ACTif} 
holds, i.e. any ASM model. Our continuous characterization can hence be viewed as subsuming
the various discrete characterizations.

Another class of models where communication is by message passing
is called \emph{dynamic networks}. Here reliable  processes run in synchronous rounds,
an adversary defines the possible patterns of message losses. Some adversaries
define models that are ASM, and hence our results apply.
 An investigation of the minimum set of messages whose delivery must be guaranteed to ensure the equivalence of this model to asynchronous wait-free shared-memory is presented in~\cite{AFEK201588}.
For general adversaries,
 a characterization is known only for consensus~\cite{NSW19:subm}.


\bibliographystyle{IEEEtran}  
\bibliography{references}

\begin{thebibliography}{10}
\providecommand{\url}[1]{#1}
\csname url@samestyle\endcsname
\providecommand{\newblock}{\relax}
\providecommand{\bibinfo}[2]{#2}
\providecommand{\BIBentrySTDinterwordspacing}{\spaceskip=0pt\relax}
\providecommand{\BIBentryALTinterwordstretchfactor}{4}
\providecommand{\BIBentryALTinterwordspacing}{\spaceskip=\fontdimen2\font plus
\BIBentryALTinterwordstretchfactor\fontdimen3\font minus
  \fontdimen4\font\relax}
\providecommand{\BIBforeignlanguage}[2]{{%
\expandafter\ifx\csname l@#1\endcsname\relax
\typeout{** WARNING: IEEEtran.bst: No hyphenation pattern has been}%
\typeout{** loaded for the language `#1'. Using the pattern for}%
\typeout{** the default language instead.}%
\else
\language=\csname l@#1\endcsname
\fi
#2}}
\providecommand{\BIBdecl}{\relax}
\BIBdecl

\bibitem{HS99:ACT}
\BIBentryALTinterwordspacing
M.~Herlihy and N.~Shavit, ``The topological structure of asynchronous
  computability,'' \emph{J. ACM}, vol.~46, no.~6, pp. 858--923, Nov. 1999,
  conference version in ACM STOC 1993. [Online]. Available:
  \url{http://doi.acm.org/10.1145/331524.331529}
\BIBentrySTDinterwordspacing

\bibitem{HoestS06}
\BIBentryALTinterwordspacing
G.~Hoest and N.~Shavit, ``Toward a topological characterization of asynchronous
  complexity,'' \emph{{SIAM} J. Comput.}, vol.~36, no.~2, pp. 457--497, 2006.
  [Online]. Available: \url{https://doi.org/10.1137/S0097539701397412}
\BIBentrySTDinterwordspacing

\bibitem{AttiyaR02}
\BIBentryALTinterwordspacing
H.~Attiya and S.~Rajsbaum, ``The combinatorial structure of wait-free solvable
  tasks,'' \emph{{SIAM} J. Comput.}, vol.~31, no.~4, pp. 1286--1313, 2002.
  [Online]. Available: \url{https://doi.org/10.1137/S0097539797330689}
\BIBentrySTDinterwordspacing

\bibitem{BG93:STOC}
E.~Borowsky and E.~Gafni, ``Generalized {FLP} impossibility result for
  t-resilient asynchronous computations,'' in \emph{STOC '93: Proceedings of
  the twenty-fifth annual ACM symposium on Theory of computing}.\hskip 1em plus
  0.5em minus 0.4em\relax New York, NY, USA: ACM, 1993, pp. 91--100.

\bibitem{SZ00:SIAM}
M.~Saks and F.~Zaharoglou, ``Wait-free k-set agreement is impossible: The
  topology of public knowledge,'' \emph{SIAM J. Comput.}, vol.~29, no.~5, pp.
  1449--1483, 2000.

\bibitem{FLP85}
M.~J. Fischer, N.~A. Lynch, and M.~S. Paterson, ``Impossibility of distributed
  consensus with one faulty process,'' \emph{Journal of the ACM}, vol.~32,
  no.~2, pp. 374--382, Apr. 1985.

\bibitem{Castaeda2010NewCT}
A.~Casta{\~n}eda and S.~Rajsbaum, ``New combinatorial topology bounds for
  renaming: the lower bound,'' \emph{Distributed Computing}, vol.~22, pp.
  287--301, 2010.

\bibitem{AlcantaraCFR19}
\BIBentryALTinterwordspacing
M.~Alcantara, A.~Casta{\~{n}}eda, D.~Flores{-}Pe{\~{n}}aloza, and S.~Rajsbaum,
  ``The topology of look-compute-move robot wait-free algorithms with hard
  termination,'' \emph{Distributed Comput.}, vol.~32, no.~3, pp. 235--255,
  2019. [Online]. Available: \url{https://doi.org/10.1007/s00446-018-0345-3}
\BIBentrySTDinterwordspacing

\bibitem{HKR13}
\BIBentryALTinterwordspacing
M.~Herlihy, D.~N. Kozlov, and S.~Rajsbaum, \emph{Distributed Computing Through
  Combinatorial Topology}.\hskip 1em plus 0.5em minus 0.4em\relax Morgan
  Kaufmann, 2013. [Online]. Available:
  \url{https://store.elsevier.com/product.jsp?isbn=9780124045781}
\BIBentrySTDinterwordspacing

\bibitem{GKM14:PODC}
\BIBentryALTinterwordspacing
E.~Gafni, P.~Kuznetsov, and C.~Manolescu, ``A generalized asynchronous
  computability theorem,'' in \emph{Proceedings of the 2014 ACM Symposium on
  Principles of Distributed Computing}, ser. PODC '14.\hskip 1em plus 0.5em
  minus 0.4em\relax New York, NY, USA: Association for Computing Machinery,
  2014, p. 222–231. [Online]. Available:
  \url{https://doi.org/10.1145/2611462.2611477}
\BIBentrySTDinterwordspacing

\bibitem{simpComp}
\BIBentryALTinterwordspacing
M.~Hazewinkel, \emph{Encyclopedia of Mathematics}.\hskip 1em plus 0.5em minus
  0.4em\relax EMS Press, 2018, ch. Simplicial complex. [Online]. Available:
  \url{http://encyclopediaofmath.org/index.php?title=Simplicial_complex&oldid=42757}
\BIBentrySTDinterwordspacing

\bibitem{SaraphHG18}
\BIBentryALTinterwordspacing
V.~Saraph, M.~Herlihy, and E.~Gafni, ``An algorithmic approach to the
  asynchronous computability theorem,'' \emph{J. Appl. Comput. Topol.}, vol.~1,
  no. 3-4, pp. 451--474, 2018. [Online]. Available:
  \url{https://doi.org/10.1007/s41468-018-0014-4}
\BIBentrySTDinterwordspacing

\bibitem{FNS18:PODC}
\BIBentryALTinterwordspacing
M.~F\"{u}gger, T.~Nowak, and M.~Schwarz, ``Tight bounds for asymptotic and
  approximate consensus,'' in \emph{Proceedings of the 2018 ACM Symposium on
  Principles of Distributed Computing}, ser. PODC '18.\hskip 1em plus 0.5em
  minus 0.4em\relax New York, NY, USA: ACM, 2018, pp. 325--334. [Online].
  Available: \url{http://doi.acm.org/10.1145/3212734.3212762}
\BIBentrySTDinterwordspacing

\bibitem{Kuznetsov12}
\BIBentryALTinterwordspacing
P.~Kuznetsov, ``Understanding non-uniform failure models,'' \emph{Bull.
  {EATCS}}, vol. 106, pp. 53--77, 2012. [Online]. Available:
  \url{http://eatcs.org/beatcs/index.php/beatcs/article/view/80}
\BIBentrySTDinterwordspacing

\bibitem{GR06:TCJ}
\BIBentryALTinterwordspacing
R.~Guerraoui and M.~Raynal, ``{The Alpha of Indulgent Consensus},'' \emph{The
  Computer Journal}, vol.~50, no.~1, pp. 53--67, 08 2006. [Online]. Available:
  \url{https://doi.org/10.1093/comjnl/bxl046}
\BIBentrySTDinterwordspacing

\bibitem{ACR21:OPODIS}
\BIBentryALTinterwordspacing
H.~Attiya, A.~Casta{\~n}eda, and S.~Rajsbaum, ``{Locally Solvable Tasks and the
  Limitations of Valency Arguments},'' in \emph{24th International Conference
  on Principles of Distributed Systems (OPODIS 2020)}, ser. Leibniz
  International Proceedings in Informatics (LIPIcs), Q.~Bramas, R.~Oshman, and
  P.~Romano, Eds., vol. 184.\hskip 1em plus 0.5em minus 0.4em\relax Dagstuhl,
  Germany: Schloss Dagstuhl--Leibniz-Zentrum f{\"u}r Informatik, 2021, pp.
  18:1--18:16. [Online]. Available:
  \url{https://drops.dagstuhl.de/opus/volltexte/2021/13503}
\BIBentrySTDinterwordspacing

\bibitem{BGsimple97}
\BIBentryALTinterwordspacing
E.~Borowsky and E.~Gafni, ``A simple algorithmically reasoned characterization
  of wait-free computation (extended abstract),'' in \emph{Proceedings of the
  Sixteenth Annual ACM Symposium on Principles of Distributed Computing}, ser.
  PODC '97.\hskip 1em plus 0.5em minus 0.4em\relax New York, NY, USA:
  Association for Computing Machinery, 1997, p. 189?198. [Online]. Available:
  \url{https://doi.org/10.1145/259380.259439}
\BIBentrySTDinterwordspacing

\bibitem{GafniR10simul}
\BIBentryALTinterwordspacing
E.~Gafni and S.~Rajsbaum, ``Distributed programming with tasks,'' in
  \emph{Principles of Distributed Systems - 14th International Conference,
  {OPODIS} 2010, Tozeur, Tunisia, December 14-17, 2010. Proceedings}, ser.
  Lecture Notes in Computer Science, C.~Lu, T.~Masuzawa, and M.~Mosbah, Eds.,
  vol. 6490.\hskip 1em plus 0.5em minus 0.4em\relax Springer, 2010, pp.
  205--218. [Online]. Available:
  \url{https://doi.org/10.1007/978-3-642-17653-1\_17}
\BIBentrySTDinterwordspacing

\bibitem{BGLR01}
E.~Borowsky, E.~Gafni, N.~Lynch, and S.~Rajsbaum, ``The {BG} distributed
  simulation algorithm,'' \emph{Distributed Computing}, vol.~14, no.~3, pp.
  127--146, 2001.

\bibitem{extendedBG09}
\BIBentryALTinterwordspacing
E.~Gafni, ``The extended bg-simulation and the characterization of
  t-resiliency,'' in \emph{Proceedings of the Forty-First Annual ACM Symposium
  on Theory of Computing}, ser. STOC '09.\hskip 1em plus 0.5em minus
  0.4em\relax New York, NY, USA: Association for Computing Machinery, 2009, p.
  85–92. [Online]. Available: \url{https://doi.org/10.1145/1536414.1536428}
\BIBentrySTDinterwordspacing

\bibitem{petr2020}
P.~Kuznetsov and T.~Rieutord, ``Affine tasks for k-test-and-set,'' in
  \emph{Stabilization, Safety, and Security of Distributed Systems},
  S.~Devismes and N.~Mittal, Eds.\hskip 1em plus 0.5em minus 0.4em\relax Cham:
  Springer International Publishing, 2020, pp. 151--166.

\bibitem{HerlihyR12simul}
\BIBentryALTinterwordspacing
M.~Herlihy and S.~Rajsbaum, ``Simulations and reductions for colorless tasks,''
  in \emph{{ACM} Symposium on Principles of Distributed Computing, {PODC} '12,
  Funchal, Madeira, Portugal, July 16-18, 2012}, D.~Kowalski and A.~Panconesi,
  Eds.\hskip 1em plus 0.5em minus 0.4em\relax {ACM}, 2012, pp. 253--260.
  [Online]. Available: \url{https://doi.org/10.1145/2332432.2332483}
\BIBentrySTDinterwordspacing

\bibitem{HRRS17:TCS}
\BIBentryALTinterwordspacing
M.~Herlihy, S.~Rajsbaum, M.~Raynal, and J.~Stainer, ``From wait-free to
  arbitrary concurrent solo executions in colorless distributed computing,''
  \emph{Theoretical Computer Science}, vol. 683, pp. 1 -- 21, 2017. [Online].
  Available:
  \url{http://www.sciencedirect.com/science/article/pii/S0304397517303298}
\BIBentrySTDinterwordspacing

\bibitem{Gafni1999ThreeProcessorTA}
E.~Gafni and E.~Koutsoupias, ``Three-processor tasks are undecidable,''
  \emph{SIAM J. Comput.}, vol.~28, pp. 970--983, 1999.

\bibitem{decidHR97}
\BIBentryALTinterwordspacing
M.~Herlihy and S.~Rajsbaum, ``The decidability of distributed decision tasks
  (extended abstract),'' in \emph{Proceedings of the Twenty-Ninth Annual ACM
  Symposium on Theory of Computing}, ser. STOC '97.\hskip 1em plus 0.5em minus
  0.4em\relax New York, NY, USA: Association for Computing Machinery, 1997, p.
  589–598. [Online]. Available: \url{https://doi.org/10.1145/258533.258652}
\BIBentrySTDinterwordspacing

\bibitem{SaraphHG16}
\BIBentryALTinterwordspacing
V.~Saraph, M.~Herlihy, and E.~Gafni, ``Asynchronous computability theorems for
  t-resilient systems,'' in \emph{Distributed Computing - 30th International
  Symposium, {DISC} 2016, Paris, France, September 27-29, 2016. Proceedings},
  ser. Lecture Notes in Computer Science, C.~Gavoille and D.~Ilcinkas, Eds.,
  vol. 9888.\hskip 1em plus 0.5em minus 0.4em\relax Springer, 2016, pp.
  428--441. [Online]. Available:
  \url{https://doi.org/10.1007/978-3-662-53426-7\_31}
\BIBentrySTDinterwordspacing

\bibitem{HerlihyR13topAdv}
\BIBentryALTinterwordspacing
M.~Herlihy and S.~Rajsbaum, ``The topology of distributed adversaries,''
  \emph{Distributed Comput.}, vol.~26, no.~3, pp. 173--192, 2013. [Online].
  Available: \url{https://doi.org/10.1007/s00446-013-0189-9}
\BIBentrySTDinterwordspacing

\bibitem{KRH18:PODC}
\BIBentryALTinterwordspacing
P.~Kuznetsov, T.~Rieutord, and Y.~He, ``An asynchronous computability theorem
  for fair adversaries,'' in \emph{Proceedings of the 2018 {ACM} Symposium on
  Principles of Distributed Computing, {PODC} 2018, Egham, United Kingdom, July
  23-27, 2018}, 2018, pp. 387--396. [Online]. Available:
  \url{https://dl.acm.org/citation.cfm?id=3212765}
\BIBentrySTDinterwordspacing

\bibitem{Delporte-Gallet18}
\BIBentryALTinterwordspacing
C.~Delporte{-}Gallet, H.~Fauconnier, S.~Rajsbaum, and N.~Yanagisawa, ``A
  characterization of t-resilient colorless task anonymous solvability,'' in
  \emph{Structural Information and Communication Complexity - 25th
  International Colloquium, {SIROCCO} 2018, Ma'ale HaHamisha, Israel, June
  18-21, 2018, Revised Selected Papers}, ser. Lecture Notes in Computer
  Science, Z.~Lotker and B.~Patt{-}Shamir, Eds., vol. 11085.\hskip 1em plus
  0.5em minus 0.4em\relax Springer, 2018, pp. 178--192. [Online]. Available:
  \url{https://doi.org/10.1007/978-3-030-01325-7\_18}
\BIBentrySTDinterwordspacing

\bibitem{HerlihyR10TopAdvPODC}
\BIBentryALTinterwordspacing
M.~Herlihy and S.~Rajsbaum, ``The topology of shared-memory adversaries,'' in
  \emph{Proceedings of the 29th Annual {ACM} Symposium on Principles of
  Distributed Computing, {PODC} 2010, Zurich, Switzerland, July 25-28, 2010},
  A.~W. Richa and R.~Guerraoui, Eds.\hskip 1em plus 0.5em minus 0.4em\relax
  {ACM}, 2010, pp. 105--113. [Online]. Available:
  \url{https://doi.org/10.1145/1835698.1835724}
\BIBentrySTDinterwordspacing

\bibitem{GWSR19:SSS}
\BIBentryALTinterwordspacing
H.~R. Galeana, K.~Winkler, U.~Schmid, and S.~Rajsbaum, ``A topological view of
  partitioning arguments: Reducing k-set agreement to consensus,'' in
  \emph{Stabilization, Safety, and Security of Distributed Systems - 21st
  International Symposium, {SSS} 2019, Pisa, Italy, October 22-25, 2019,
  Proceedings}, ser. Lecture Notes in Computer Science, vol. 11914.\hskip 1em
  plus 0.5em minus 0.4em\relax Springer, 2019, pp. 307--322. [Online].
  Available: \url{https://doi.org/10.1007/978-3-030-34992-9\_25}
\BIBentrySTDinterwordspacing

\bibitem{soloACT2020}
\BIBentryALTinterwordspacing
Y.~Yue, F.~Lei, X.~Liu, and J.~Wu, ``Asynchronous computability theorem in
  arbitrary solo models,'' \emph{Mathematics}, vol.~8, no.~5, 2020. [Online].
  Available: \url{https://www.mdpi.com/2227-7390/8/5/757}
\BIBentrySTDinterwordspacing

\bibitem{BenavidesR18}
\BIBentryALTinterwordspacing
F.~Benavides and S.~Rajsbaum, ``Collapsibility of read/write models using
  discrete morse theory,'' \emph{J. Appl. Comput. Topol.}, vol.~1, no. 3-4, pp.
  365--396, 2018. [Online]. Available:
  \url{https://doi.org/10.1007/s41468-018-0011-7}
\BIBentrySTDinterwordspacing

\bibitem{AFEK201588}
\BIBentryALTinterwordspacing
Y.~Afek and E.~Gafni, ``A simple characterization of asynchronous
  computations,'' \emph{Theoretical Computer Science}, vol. 561, pp. 88--95,
  2015, special Issue on Distributed Computing and Networking. [Online].
  Available:
  \url{https://www.sciencedirect.com/science/article/pii/S0304397514005659}
\BIBentrySTDinterwordspacing

\bibitem{NSW19:subm}
T.~Nowak, U.~Schmid, and K.~Winkler, ``Topological characterization of
  consensus under general message adversaries,'' 2019, (submitted to PODC'19).

\end{thebibliography}

\end{document}